\documentclass[fleqn,usenatbib]{mnras}
\usepackage{newtxtext,newtxmath}
\usepackage[T1]{fontenc}
\usepackage{ulem}
\DeclareRobustCommand{\VAN}[3]{#2}
\let\VANthebibliography\thebibliography
\def\thebibliography{\DeclareRobustCommand{\VAN}[3]{##3}\VANthebibliography}

\usepackage{graphicx}	
\usepackage{amsmath}	
\usepackage{comment}

\usepackage{float}

\usepackage{graphicx} 
\usepackage[export]{adjustbox}
\usepackage{caption,subcaption}

\usepackage{tikz,xcolor,hyperref}
\definecolor{lime}{HTML}{A6CE39}
\DeclareRobustCommand{\orcidicon}{%
	\begin{tikzpicture}
	\draw[lime, fill=lime] (0,0) 
	circle [radius=0.16] 
	node[white] {{\fontfamily{qag}\selectfont \tiny ID}};
	\draw[white, fill=white] (-0.0625,0.095) 
	circle [radius=0.007];
	\end{tikzpicture}
	\hspace{-2mm}
}
\foreach \x in {A, ..., Z}{%
	\expandafter\xdef\csname orcid\x\endcsname{\noexpand\href{https://orcid.org/\csname orcidauthor\x\endcsname}{\noexpand\orcidicon}}
}

\author[Mondal, Pramanick, Resmi \& Bose]{
Tanima Mondal\orcidB,$^{1}$\thanks{mtanima14@gmail.com}
Suman Pramanick\orcidA,$^{1}$\thanks{suman21eor@gmail.com}
Lekshmi Resmi\orcidC$^{2}$\thanks{l.resmi@gmail.com }
and Debanjan Bose\orcidD$^{3}$\thanks{debaice@gmail.com}
\\
$^{1}$Department of Physics, Indian Institute of Technology Kharagpur, Kharagpur, West Bengal 721302 India\\
$^{2}$Department of Earth \& Space Sciences, Indian Institute of Space Science \& Technology, Trivandrum 695547, India\\
$^{3}$School of Astrophysics, Presidency University, Kolkata 700073, India
}

\date{Accepted XXX. Received YYY; in original form ZZZ}

\pubyear{2021}

\begin{document}
\label{firstpage}
\pagerange{\pageref{firstpage}--\pageref{lastpage}}
\title[GRB Afterglows with CTA]{Probing Gamma-Ray Burst afterglows with the Cherenkov Telescope Array}
\maketitle

\begin{abstract}
Detection of delayed sub-TeV photons from Gamma-Ray Bursts (GRBs) by MAGIC and HESS has proven the promising future of GRB afterglow studies with the Cherenkov Telescope Array (CTA), the next-generation gamma-ray observatory. With the unprecedented sensitivity of CTA, afterglow detection rates are expected to increase dramatically. In this paper, we explore the multi-dimensional afterglow parameter space to see the detectability of sub-TeV photons by CTA. We use a one-zone electron synchrotron and synchrotron self-Compton model to obtain the spectral energy distribution. We consider bursts going off in a medium of homogenous density. The blast wave is assumed to be radiatively inefficient and evolving adiabatically. Considering that the electron acceleration is not efficient if the acceleration timescale exceeds the radiative cooling timescale, we find that the Sub-TeV emission is always due to the self-Compton process. We find that jets with high kinetic energy or large bulk Lorentz factor decelerating into a dense ambient medium offer better detection prospects for CTA. For relatively lower values of the downstream magnetic field, electrons are slow-cooling, and the emitted radiation is positively correlated with the magnetic field. For larger magnetic fields, the electron population enters the fast cooling phase where the radiated flux is inversely proportional to the magnetic field.  We apply our results in the context of bright TeV afterglows detected in recent years. Our results indicate that cosmological short GRBs have only moderate prospects of detection by CTA while local Neutron Star merger counterparts can be detected if the jet is launched towards the observer.  

\end{abstract}


\section{Introduction}

Gamma-ray Bursts (GRBs) release a large amount of energy ($10^{49}-10^{54}$ erg) in a very short time, known as prompt emission\citep{klebesadel1973observations, Costa:1997obd, Frontera:1997ae, 1997Natur.386..686V, Kulkarni:1999aa}. Thus they are the most powerful explosions in the Universe. GRBs which last for less than 2 seconds are classified as short GRBs, believed to be due to the merger of a binary compact object system involving two Neutron Stars or a Neutron Star and a Black Hole \citep{1989Natur.340..126E, 1991AcA....41..257P,duncan1992formation, usov1992stellar,thompson1994model,metzger2011protomagnetar}. Long GRBs, which last for more than 2 seconds, are believed to be due to the gravitational collapse at the end of a massive star \citep{woosley1993gamma, paczynski1998gamma, macfadyen1999collapsars, woosley2006supernova}. 

A large amount of energy is released predominantly in $\gamma$-rays due to dissipative processes inside the ejected material during the prompt emission phase. According to the standard fireball model \citep{meszaros1992high,meszaros1993relativistic,piran1993hydrodynamics, piran1993fireballs,granot1999images}
, radiation pressure overcomes gravity and heats matter into a fireball, which then expands relativistically and drives a blast wave through the ambient medium. Afterglow radiation emerges from the kinetic energy of the ejecta dissipated in this medium. Afterglow spectral energy distribution (SED) extends from radio to $\gamma$-ray frequencies. Both prompt emission and afterglow arise from non-thermal radiative processes of high-energy particles. See \cite{kumar2015physics} or  \cite{2018pgrb.book.....Z} and references therein for a detailed description of the standard afterglow model explaining both the prompt and afterglow emission.

Currently, a broad picture of the physics of GRBs is established fairly well. However, many finer details, such as the generation of magnetic field and acceleration of high-energy particles by the relativistic shock in its downstream, are still to be understood \citep{2009MNRAS.400L..75K, Miceli:2022efx, Gill:2022erf}. In addition, the basic model has failed to reproduce the observed spectral evolution in some well-observed afterglows  \citep{fraija2022modeling, Misra2021, Rhodes2020}. The afterglow phenomena need to be probed in novel ways to fill the gaps in our understanding. Most ($\sim 95$\%) GRBs have well-sampled X-ray afterglow lightcurves, thanks to the fast slewing ability of the Neil Gehrels \textit{Swift} observatory\footnote{\url{https://swift.gsfc.nasa.gov/}}. $\gamma$-ray afterglows started to get discovered by the launch of {\textit {Fermi}} satellite\footnote{\url{https://fermi.gsfc.nasa.gov/}}, where the Large Area Telescope (LAT)\footnote{\url{https://glast.sites.stanford.edu/}} started detecting delayed GeV photons in many bursts 
\citep{ajello2019decade,2014ApJ...788..156T, 2013ApJ...773L..20L}. 

A novel window of afterglow observations opened up with the recent detections of GeV-TeV photons arriving in timescale of a few minutes since burst, indicating their afterglow origin. Sub-TeV photons from six GRBs\footnote{\url{http://tevcat.uchicago.edu/}} are detected by ground-based Cherenkov telescopes in the last 5 years \citep{Abdalla:2019dlr,veres2019observation, MAGIC:2019lau, HESS:2021dbz, 2020GCN.29075....1B, 2022GCN.32677....1H}. The rates will improve with the upcoming Cherenkov Telescope Array (CTA)\footnote{\url{https://www.cta-observatory.org/}} \citep{acharya2017science}, and will be opening an unprecedented opportunity for GRB physics. 

Afterglow emission arises from non-thermal electrons accelerated by relativistic shocks. These electrons will radiate via synchrotron and Synchrotron Self Compton (SSC) processes. A majority of afterglows observed in radio to X-ray frequencies are due to synchrotron emission by these electrons. For most of the standard parameters, SSC photons start to dominate beyond X-ray or GeV frequencies  \citep{fraija2019synchrotron,fraija2021origin, 2022ApJ...934..188F,2013ApJ...776...95F, 2019ApJ...884..117W}. This means that the SSC component has not been studied observationally as extensively as the synchrotron component  until now. 

With the LAT and CTA covering the GeV-TeV band, we are now in a position to explore the SSC process in afterglows both observationally and theoretically \citep{2019ICRC...36..634B}. Theoretical predictions of SSC SED can now be tested against the data. No additional parameters enter the problem while modelling the SSC component as the process is self-Compton, not external Compton. Therefore, by probing the new window, a better estimate of the afterglow parameter space, especially the ones governing the shock microphysics, is possible. Since shock microphysics is difficult to predict from the first principles, several simplifying assumptions are used in the standard afterglow model. This is an opportunity to test these assumptions as well, which has already begun \citep{Misra2021, Rhodes2020, veres2019observation}.  

In this paper, we calculate the temporal and spectral evolution of GRB afterglows focusing on the sub-TeV regime to explore GRB afterglow detection probabilities in the era of the CTA. The paper is organized as follows. In section~\ref{Methods}, we have discussed our methodology on relativistic blast wave dynamics and time-evolving spectral energy distribution of synchrotron and synchrotron self-Compton (SSC) radiation to obtain VHE GRB afterglow emissions. Extinction of very high energy gamma-ray flux due to EBL (Extra-Galactic Background Light) is discussed in Section~\ref{EBL correction}. In section \ref{Flux evolution in CTA bands}, we have presented light curves and SEDs for a wide range of physical parameters which are likely to be detected by CTA. In section~\ref{Discussion}, we have shown that our model can explain the VHE afterglow emissions for GRBs detected by ground-based gamma-ray telescopes. Finally, we summarize the paper in section~\ref{conclusion}.

\section{Methods}\label{Methods}
In order to obtain the time-evolving afterglow flux from both synchrotron and SSC processes, we have to calculate the time evolution of the magnetic field and electron population downstream. In this section, we describe the dynamics of the relativistic forward shock arising from the interaction of the ejecta with the ambient medium. We have not considered the reverse shock emission in this paper though it is expected to contribute to early emission in the GeV \citep{fraija2020grb}.

We consider an ultra-relativistic radially symmetric blast-wave decelerating into a constant density ambient medium. We ignore the lateral expansion phase as it happens in the timescale of a few days for standard afterglow parameters and hence is beyond the reach of VHE telescopes. Since we restrict ourselves to the spherically symmetric initial phase, the opening angle of the jet does not enter the problem. Therefore, the basic parameters are $E$, the isotropic equivalent kinetic energy in the explosion, $n_0$ the ambient density, $p$ the power-law index of the non-thermal electron distribution, $\epsilon_e$ the fractional thermal energy density in the shock downstream transferred to the non-thermal electrons, and $\epsilon_B$ the same transferred to the magnetic field. 

The blast wave deceleration is assumed to be adiabatic, i.e., the radiative loss from downstream is considered negligible. To be self-consistent with this assumption, we have assigned a relatively lower fraction of downstream thermal energy in the electron population (see section \ref{dynamics}). 

We consider the simplistic model where the fractional energy in non-thermal electrons and magnetic fields remain constant throughout the evolution. However, there are indications for a deviation from this for TeV bright bursts, particularly \citep{Misra2021, Rhodes2020}.

\subsection{Dynamics}
\label{dynamics}
The external forward shock model is based on the interaction of ultra-relativistic ejecta with the ambient medium around the burster. Both the ejecta and the ambient medium are considered to be cold (negligible pressure and temperature) to begin with. The blast wave ahead of the ejected shell decelerates as it transfers energy and momentum to the surrounding medium. 

VHE lightcurves typically peak at the early phase of the afterglow, around the beginning of the blast wave deceleration. Therefore, to faithfully reproduce VHE lightcurves, it is important to estimate Lorentz factor evolution at the onset of deceleration accurately. Keeping this in mind, to obtain the afterglow dynamics, we solve the set of ordinary differential equations governing the evolution of the blast-wave Lorentz factor and radius, adopting the treatment by \cite{pe2012dynamical}. 

We begin with the evolution of the bulk Lorentz factor ($\Gamma$) of the blast wave as it collects material from the ISM (equation-5 in the original paper),
\begin{equation}
\label{gmeqn}
    \frac{d \Gamma}{d m} = - \frac{\hat{\gamma}(\Gamma^2 - 1) - (\hat{\gamma}-1) \Gamma \beta^2}{M + \epsilon m + (1-\epsilon) m \left[2 \hat{\gamma} \Gamma - (\hat{\gamma}-1)(1+\Gamma^{-2}) \right]},
\end{equation}
where $m$ is the rest mass of the ambient medium, $M$ is the total mass of the ejected shell, and $\epsilon$ is the fractional thermal energy radiated away from the downstream \citep{Huang:1999di, Huang:1999qa}.

The adiabatic index ($\hat{\gamma}$) of the downstream plasma is a function of the plasma temperature, which evolves with respect to $\Gamma$. As in \cite{pe2012dynamical}, we use the polynomial fit from \cite{Service1986ApJ} to calculate $\hat{\gamma}$ as a function of $\Gamma$.

The rest mass of the ambient medium $m$ is given by,
\begin{equation}
    dm = 4 \pi r^2 \rho dr, 
\label{mreqn}
\end{equation}
where $r$ is the radius of the shock front from the explosion site and $\rho$ is the density of the medium. 

The photon arrival time ($t$) for the observer is given by, 
\begin{equation}
    \frac{dt}{1+z} = dr\, \frac{(1-\beta)}{\beta\,c},
\label{treqn}
\end{equation}
where $\beta$ is the velocity corresponding to the Lorentz factor $\Gamma$, $z$ is the cosmological redshift of the source, and $c$ is the speed of light. Here we have assumed the velocity vector to be aligned towards the observer's line of sight. 

The three simultaneous ordinary differential equations above describe the evolution of the bulk Lorentz factor of the blast wave as a function of the observer's time. We solve the equations using the ODE solver from the \textit{scipy}\footnote{\url{www.scipy.org}} package using the initial condition of $r=0$, $m=0$, and $\Gamma = \Gamma_0$. The Lorentz factor decreases as the blast wave starts to decelerate. Analytical approximations consider $\Gamma \propto r^{-3/2}$ for $r>r_{\rm dec}$ and $\Gamma_0$ otherwise, where $r_{\rm dec} = 10^{18} {\rm cm} (E/10^{52} {\rm erg})^{1/3} n_0^{-1/3} \Gamma_0^{-2/3}$ is the deceleration radius at which the downstream inertial mass is comparable to the mass in the ejecta. $n_0$ is normalized to $1 \,{\rm cm}^{-3}$.

\begin{figure}
    \centering
    \includegraphics[width=0.48\textwidth]{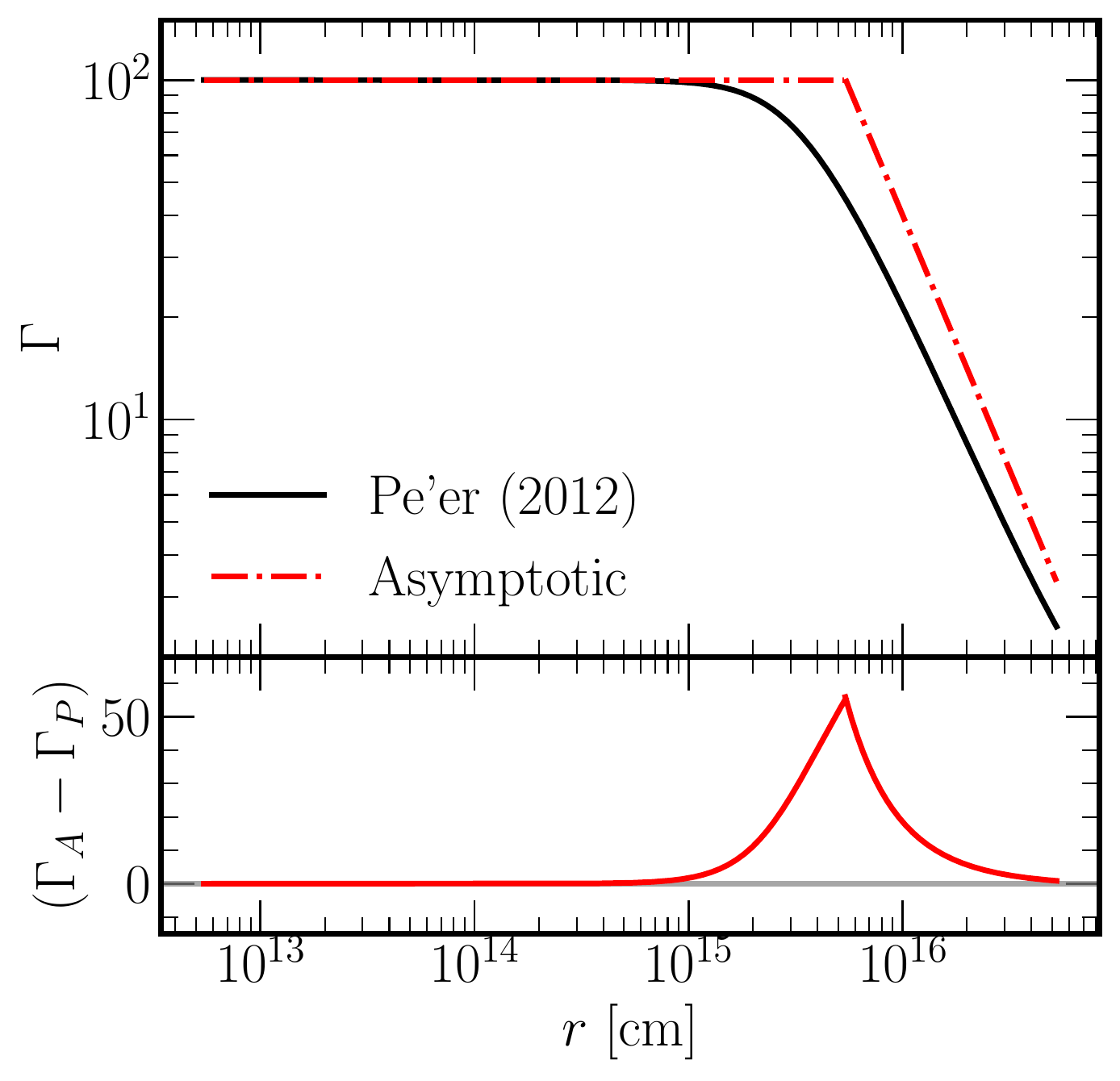}
    \caption{This shows the variation of the bulk Lorentz factor $\Gamma$ with respect to the blast wave radius $r$ for \citealt{pe2012dynamical} model (black solid line) and asymptotic solution (red dashed line). The bottom panel (red solid line) shows the difference between the asymptotic $(\Gamma_A)$ and the \citealt{pe2012dynamical} $(\Gamma_P)$ solutions. These plots are generated for $E = 10^{50}$ erg, $\Gamma_0 = 100$ and $n=10$ cm$^{-3}$. }
    \label{fig:dynamics}
\end{figure}

In figure~\ref{fig:dynamics}, we present the evolution of $\Gamma(r)$ from the numerical calculation and the analytical approximation. The difference between the two is shown in the bottom panel. The analytical approximation deviates considerably near the epoch of deceleration where the VHE light curve peaks, leading to an inaccurate estimate of the peak time and the peak flux.

Once $\Gamma(r)$ and $r(t)$ are obtained, we proceed to calculate the synchrotron and SSC flux from the downstream.

\subsection{Radiation spectrum}\label{Spectral Energy Distribution}
GRB afterglows originate from synchrotron and Synchrotron Self-Compton (SSC) radiation from the blast wave downstream. 
In this section, we describe the calculation of the time-evolving SEDs of synchrotron and SSC radiation following the framework described in \citet{Sari:1997qe, wijers1999physical, panaitescu2000analytic, Sari_2001}. The time evolution of the SED is decided by the dynamics discussed in the previous section. 

The first step in calculating the SED is to obtain the energy distribution of non-thermal electrons and the energy density of the  magnetic field in the shock downstream. 

If the electrons carry a significant fraction of the downstream thermal energy, the blast wave evolution becomes radiative when radiative losses of the electron population are significant \citep{Huang:1999qa}. In such a scenario, one has to consider the transition of the afterglow dynamics from radiative to the adiabatic regime. We assumed a relatively lower $\epsilon_e$ to avoid this complication and ensure that radiative losses from the shock downstream are negligible and that the thermal energy density remains intact. For our calculations, we fixed $\epsilon_e = 0.02$.

\subsubsection{Synchrotron spectrum}
The electrons are distributed as a power-law, specified by the minimum and maximum electron Lorentz factors ($\gamma_m$ and $\gamma_{\rm max}$ respectively), and the power-law index $p$. Using the downstream number density and thermal energy density, $\gamma_m$ and $B$ can be estimated as $(m_p/m_e) (p-2)/(p-1) \epsilon_e \Gamma$ and $(\epsilon_B 4 n_0 m_e c^2)^{1/2} \Gamma$ respectively, where $m_p$ is the rest mass of protron and $m_e$ is the rest mass of electron. We have assumed the entire downstream electrons to be distributed in the non-thermal population.  

The injected electron distribution $\propto \gamma^{-p}$ is affected due to energy loss from synchrotron radiation. The comoving timescale $t_{1/2}^{\prime}$ for an electron to lose half of its initial kinetic energy is given by $(\gamma-1) m_e c^2/P_{\rm syn}(\gamma)$, where $\gamma$ is the random Lorentz factor of the electron and $P_{\rm syn}(\gamma)$ is its total synchrotron power. For a detailed derivation, see \cite{Rybicki:2004hfl}. At a given time $t$ corresponding to the co-moving time $t^{\prime} = 2 \Gamma t$, electrons with $t_{1/2}^{\prime} \le t^{\prime}$ suffer severe synchrotron losses. The critical Lorentz factor is obtained by equating the two timescales and is given by $\gamma_c^s(t) = \frac{(1+z) 6\pi m_{e}c}{\sigma_{T} B^{2} \Gamma t}$.

Electrons above $\gamma_c^s$ deviate from the injected power law and settle into a distribution $\propto \gamma^{-(p+1)}$ asymptotically. Therefore, the final synchrotron spectrum can be approximated as piecewise power-law segments separated by $\nu_m=\nu_{\rm syn}(\gamma_m), \nu_c^s = \nu_{\rm syn}(\gamma_c^s)$ and $\nu_{\rm max} = \nu_{\rm syn}(\gamma_{\rm max}$), where $\nu_{\rm syn}(\gamma)$ is the characteristic frequency related to the peak of the single electron synchrotron power and is given by $e B \gamma^2 \Gamma/(2 \pi m_e c$). The flux $f_{\nu} \propto \nu^{1/3}$ if $\nu < \nu_m$, $\nu^{(-p-1)/2}$ if $\nu_m< \nu< \nu_c^s$, and $\nu^{-(p/2)}$ if $\nu_c^s < \nu$. This is valid only for the slow cooling phase ($\gamma_m < \gamma_c^{s}$.)

If all the electrons in the distribution suffer synchrotron losses (fast cooling phase), the distribution function changes to $\propto \gamma^{-2}$ for $\gamma_c^{s} \le \gamma < \gamma_m$ and $\propto %
\gamma^{-(p+1)}$ above $\gamma_m$, leading to $f_{\nu} \propto \nu^{1/3}$ for $\nu < \nu_c^{s}$, %
$\nu^{-1/2}$ if $\nu_c^s< \nu< \nu_m$, and $\nu^{-(p/2)}$ if $\nu_m < \nu$.

Electrons can not be accelerated efficiently to Lorentz factors for which the radiative cooling timescale is shorter than the acceleration timescale \citep{gallant1999particle, achterberg2001particle, 2009herb.book.....D}. Therefore, the maximum synchrotron energy can be derived by equating the two timescales. The acceleration timescale is decided by electron confinement in the shock, which in turn is decided by the gyration radius $r_B = \gamma m_e c^2/e \, B$ for an electron of Lorentz factor $\gamma$ (here, $e$ is the charge of the electron).

Hence, the acceleration timescale in the co-moving frame, $t_{\rm acc}^{\prime} =r_B/(e_{\rm acc} \, c) = \frac{\gamma m_e c}{e_{\rm acc} e B}$ \citep{achterberg2001particle, 10.1111/j.1365-2966.2010.17927.x, 2018pgrb.book.....Z}. $e_{\rm acc}$ is a factor of the order of unity, specific to the acceleration process representing its efficiency.

Comparing $t_{\rm acc}^{\prime}$ with the synchrotron cooling time, $t_{1/2}^{\prime} = 6 \pi m_e c/\sigma_T B^2 \gamma$ (for $\gamma>$1), we can derive the maximum synchrotron energy $\gamma_{\rm max}$ as $\sqrt{6 \pi e\, e_{\rm acc} /\sigma_T B}$. We fix $e_{\rm acc}=0.35$ in our calculations \citep{zhang2020inverse}).

\subsubsection{Synchrotron Self-Compton spectrum}  
In order to obtain the radiated power in SSC, first the Compton-Y parameter that relates the seed synchrotron photon field to the upscattered radiation field needs to be calculated. In the regime of Thomson scattering, Y-parameter does not depend on the electron or photon energy, and an asymptotic solution can be given as \citep{Jacovich:2020teu}, 
\begin{equation*}
Y=
    \begin{cases}
        \left ( \frac{\epsilon_{e}}{\epsilon_{B}} \frac{1}{3-p}\left ( \frac{\gamma_{m}}{\gamma _{c}^{s}} \right )^{p-2}\right )^{\frac{1}{4-p}} & Y>>1,\\ 
        \frac{\epsilon_{e}}{\epsilon_{B}} \frac{1}{3-p}\left ( \frac{\gamma_{m}}{\gamma _{c}^{s}} \right )^{p-2} & Y<<1, 
    \end{cases}
\end{equation*}
for the slow cooling phase.
And for the fast cooling phase,
\begin{equation*}
Y=
\begin{cases}
     \left ( \frac{\epsilon_{e}}{\epsilon_{B}} \frac{1}{3-p}\right )^{\frac{1}{4-p}} & Y>>1, \\
     \left ( \frac{\epsilon_{e}}{\epsilon_{B}} \frac{1}{3-p}\right ) & Y<<1.
\end{cases}
\end{equation*}

The Y-parameter is also relevant to estimate the cooling Lorentz factor in cases where the total SSC power is greater than the synchrotron power. After including SSC losses, the effective cooling Lorentz factor is modified as $\gamma_{c}=\gamma_{c}^{s}/(1+Y)$. Hence the synchrotron cooling frequency reduces to $\nu_{c}=\nu_c^{s}/(1+Y)^{2}$. 

In this paper, we have considered inverse Compton scattering in the Thomson regime, where the scattering is elastic, and the cross-section is insensitive to the energy of the electron or the photon. However, the Klein-Nishina (KN) effect that reduces the scattering cross-section can be significant for photons in the GeV-TeV regime \citep{Fan:2007tqx, 2009ApJ...703..675N, 2010ApJ...712.1232W, Jacovich:2020teu}. 

At the time of deceleration, the frequency $\nu_{\rm KN}$ above which KN effect becomes significant is given by \citep{Ando:2008pj},
\begin{align*}
    \nu_{\rm KN} &= 41{\rm GeV}  \, \left(\frac{\Gamma_0}{100} \right)^2 \frac{\epsilon_e}{0.02} & \gamma_c < \gamma_m, \\
    \nu_{\rm KN} & =3 {\rm TeV} \left( \frac{\Gamma_0}{100} \right)^{2/3} n_0^{-2/3} \left(\frac{\epsilon_B}{10^{-3}}\right)^{-1} (1+Y)^{-1} & \gamma_m < \gamma_c.
\end{align*}

Since most of our analysis is centred around the peak of the $500$~GeV lightcurve, we compared $\nu_{\rm KN}(t_{\rm dec})$ with $500$~GeV. For a part of our parameter space where both $\epsilon_B$ and $n_0$ are large (say, $\gtrsim 10^{-2}$ and $\gtrsim 1.0$ respectively), $\nu_{\rm KN}$ can go above $500$~GeV.  This will reduce the scattering cross-section and therefore decrease the $500$~GeV flux. For frequencies $10$ times above $\nu_{\rm KN}$, the scattering cross-section drops by a factor of 10 from $\sigma_T$ and for $\nu = 100 \nu_{\rm KN}$, the drop is $\sim 50$. 
  
The characteristic frequencies for the SSC emission are $\nu_{m}^{\rm ssc} = 4\gamma_m^2 \nu_m$ and $\nu_{c}^{\rm ssc} = 4\gamma_c^2 \nu_c$. We have calculated the SSC spectrum following the prescription \cite{2018pgrb.book.....Z}, which is an adaptation of \cite{Sari_2001}. 

We have ignored optical depth to pair production in obtaining the SSC spectrum. We have also not considered the effect of $\gamma_{\rm max}$ as the corresponding frequencies are well above TeV.

In low frequencies, such as in the radio band, synchrotron self-absorption becomes relevant. Though radio photons can be upscattered to $\gamma$-ray frequencies, the corresponding flux is too low and often below the synchrotron flux for typical parameters. Therefore, we have not considered synchrotron self-absorption in our calculation and treated the SED to be optically thin always. 

\section{EBL correction}\label{EBL correction}
Extragalactic Background Light (EBL) is the diffuse electromagnetic radiation that contributes mainly to radiation from stellar nucleosynthesis and radiation from dust after absorbing starlight and AGNs. Radiation from starlight and hot dust mainly contributes to the optical, near-infrared (NIR) and far-infrared (FIR) bands. At the same time, AGNs and quasars are expected to contribute in the mid-IR band no more than 5 -- 20\% of the total EBL photon density \citep{matute2006active}. Extragalactic backgrounds of other energy bands like radio, UV, X-rays and $\gamma$-rays are one to three orders of magnitude smaller than the optical and IR backgrounds \citep{dole2006cosmic}.

The VHE $\gamma$-ray photons, which are the main observable frequencies of the ground-based atmospheric Cherenkov telescopes, undergo energy-dependent attenuation due to EBL. VHE $\gamma$-ray photons originated from extragalactic sources interact with EBL through pair production \citep{Stecker1992TeV, ackermann338allafort}. This affects the VHE part of the GRB spectrum and changes the shape of the spectrum significantly beyond those energies. A considerable amount of effort has been invested in modelling the EBL attenuation as a function of the energy of the photon and redshift \citep{franceschini2008extragalactic, finke2010modeling, dominguez2011extragalactic}. These models have been successfully used by several highly sensitive Imaging Atmospheric Cherenkov Telescopes (IACTs) such as MAGIC\footnote{\url{http://www.magic.iac.es/}} \citep{cortina2005status}, HESS\footnote{\url{https://www.mpi-hd.mpg.de/hfm/HESS/pages/about/telescopes/}} \citep{hinton2004new} and VERITAS\footnote{\url{https://veritas.sao.arizona.edu/}} \citep{holder2009status}, to analyze the observed VHE GRB afterglows from sources of different energies and situated at different redshifts. 

The observed flux can be written in terms of the intrinsic flux and the survival probability of the VHE photons as \citep{hauser2001cosmic}

\begin{equation}
    \left(\frac{d F}{d E}\right)_{\rm obs} = \left(\frac{d F}{d E}\right)_{\rm int} \times e^{-\alpha \tau(E,z)},
\end{equation}
where $\left(dF/dE\right)_{\rm int}$ is the intrinsic spectrum of the source, $\tau(E,z)$ is the optical depth of the EBL absorption, which is a function of the energy $E$ and redshift $z$. 

In this paper, we have used the EBL model of \cite{dominguez2011extragalactic} and estimate the EBL attenuation factor using a publicly available package {\texttt JetSeT} \footnote{Available at: \url{https://jetset.readthedocs.io/en/1.1.2/}} \citep{massaro2006log, tramacere2009swift,tramacere2011stochastic,tramacere2020jetset}.
In the next section, we obtain the EBL-corrected afterglow spectrum and light curves in the CTA frequencies.

\section{Flux evolution in CTA bands}\label{Flux evolution in CTA bands} 
The current statistics of TeV afterglow detections show a diversity in the afterglow behaviour. Therefore, in order to obtain the detectability of GRB afterglows in the GeV-TeV regime, it is important to explore the physical parameter space extensively. See \cite{2008A&A...489.1073G, 2009ApJ...703...60X, 2017ApJ...844...92F, 2017ApJ...846..152V, 2020MNRAS.496..974Z, 2021MNRAS.505.1718J} for some of the recent explorations in this direction.  

The peak flux and the time of peak vary from burst to burst depending on the physical parameters. In addition to the brightness of the afterglow, the timescale in which the afterglow flux remains above the threshold of the detector is also important. For example, the afterglow was detectable for timescales of a few hours in GRB1909114C \citep{veres2019observation} while it remained above the detection limit for days in the case of GRB190829A \citep{HESS:2021dbz}. 

\subsection{TeV light curve}
We consider $500$~GeV as a representative frequency to explore the afterglow behaviour in TeV ranges. The reason behind selecting this particular frequency is that it falls in the core energy range of CTA. The sensitivity of CTA is expected to be best in the energy range 100 GeV to 10 TeV \footnote{\url{https://www.cta-observatory.org/science/ctao-performance/}} (\cite{boserev}). 

For the entire parameter space we explored, the observed flux in $500$~GeV is from the SSC process as $\nu_{\rm syn} (\gamma_{\rm max})$ is below $500$~GeV. The deceleration epoch varies from $0.1$ to $10$~seconds depending on the value of $E, n_0$ and $\Gamma_0$. 

In figures- \ref{LC_1_all}, \ref{LC_2_all}, and \ref{LC_3_all}, we present the light curves in the $500$~GeV band for three different redshifts, $z=0.1, 0.5$, and $1.0$ respectively. The left panel is the flux at the source, and the right one is the observed flux after considering EBL attenuation. The vertical lines correspond to analytical approximation of 
$t_{\rm dec} = 2.3{\rm sec} \, E_{51}^{1/3} n_0^{-1/3}(\Gamma_{0,300})^{-8/3}$, where $E$ is normalized to $10^{51} {\rm erg}$, $n_0$ is normalized to $1$ ${\rm cm}^{-3}$, and $\Gamma_0$ is normalized to $300$. As we have used $\Gamma(t)$ from the simultaneous numerical solutions of equations~ \ref{gmeqn}, \ref{mreqn} and \ref{treqn}, the light curve peak does not exactly coincide with the analytical value of $t_{\rm dec}$. 

At a given redshift, low $E$, high $\Gamma_0$, and high $n_0$ lead to an early deceleration. Therefore an early peak of the light curve, at timescales of the order of a few seconds, can be seen from the figure. Even if detection at such early epochs may be impractical, the peak flux also goes up in case of high $E$ or high $n_0$, and the light curve remains above the detection level for a longer period. High values of these parameters are favourable for both detection and follow-up.  

At lower redshifts, as the afterglow is brighter, the flux can be detected till about 1000 seconds for most combinations of parameters. However, at the redshift of $1.0$, afterglows flux falls below the sensitivity limit after EBL correction, even if the initial bulk Lorentz factor is large.  

While the lightcurve peak depends on $E, n_0,$ and $\Gamma_0$, the peak flux depends on $\epsilon_B$ also in addition to these three parameters. The dependence on $\epsilon_B$ is visible in the SED at the time of the peak. For all lightcurves presented here, the SED is in the slow cooling regime. Therefore, a larger $\epsilon_B$ leads to a larger flux. The correlation reverses for the fast cooling regime (see below). 

Next, in order to explore the dependence of the afterglow flux on different parameters, we look at the SEDs of the afterglow.
\begin{figure}
\begin{centering}
\includegraphics[width=0.5\textwidth]{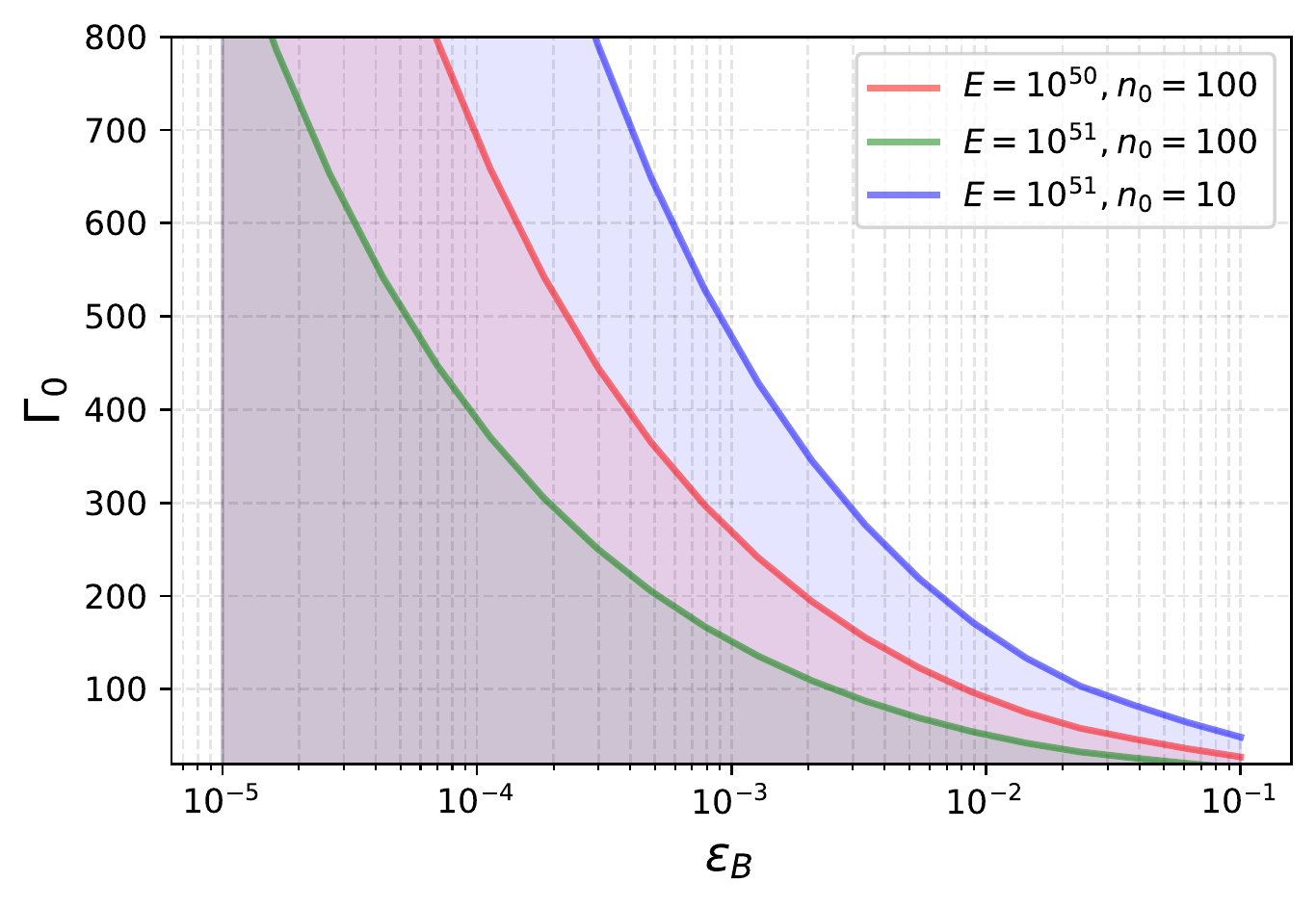}
\caption{Parameter space deciding electron cooling at $t_{\rm dec}$. The shaded region corresponds to a slow cooling electron distribution. The region above the line corresponds to fast cooling at $t_{\rm dec}$. In addition to $\epsilon_B$ and $\Gamma_0$, $n_0, E$, and $\epsilon_e$ are also important in deciding the electron cooling. We fixed $\epsilon_e = 0.02$, redshift z=0.1, and considered three different $E-n_0$ pairs in this figure.}
\label{fig2}
\par\end{centering}
\end{figure}

\subsection{Spectral energy distribution} \label{Spectral energy}

The peak flux of the light curve depends on a combination of physical parameters. The functional form is decided by the spectral regime in which $500$~GeV falls at the time of deceleration. 

First, we need to see whether the electrons are in slow or fast cooling regime at $t_{\rm dec}$. Using the analytical expression for $t_{\rm dec}$, the factor $\gamma_m/\gamma_c \propto \Gamma_0^{4/3} E^{1/3} n_0^{2/3} \epsilon_e \epsilon_B (1+Y)/(1+z)$ at the peak. Hence, for a fixed $\epsilon_e$, the most important parameters deciding between fast vs slow cooling is $\epsilon_B$ and $\Gamma_0$. In figure-\ref{fig2}, we present the $\epsilon_B - \Gamma_0$ parameter space at $t_{\rm dec}$ for three different combinations of $E$ and $n_0$. For a given $\epsilon_B$, the electrons are in the slow cooling phase if $\Gamma_0$ is below a threshold value, which is represented by solid lines in figure-\ref{fig2}. The region below the curve represents slow cooling. For a moderate $\Gamma_0$ of $300$, unless $\epsilon_B > 2 \times 10^{-4}$, the deceleration begins with the electrons in the slow cooling regime and continues to remain so because $\gamma_m$ drops faster than $\gamma_c$ post deceleration.

We present SEDs at the peak time for a few representative combinations of physical parameters in figure- \ref{SED_all}. The left panels are flux at source, and the right panels are flux corrected for EBL attenuation. The SEDs are for three redshifts, $0.1$ and $0.5$, and $1.0$. 

We fixed $\epsilon_e = 0.02$ and $p=2.2$ in these figures. About 1 GeV or below, the afterglow flux starts to get dominated by SSC. Significant EBL attenuation begins around $300$~GeV. For all cases we have presented here, the electrons are in the slow cooling regime. For all of them, $\sqrt{ {\nu_m}^{\rm SSC} \nu_c^{\rm SSC}} < 500 \,{\rm GeV} < \nu_c^{\rm SSC}$. In this spectral regime, the term $f_{\rm max}^{\rm SSC} \left( \nu/\nu_m^{\rm ssc} \right)^{-(p-1)/2}$ and a logarithmic term consisting of $\log(\nu_c^{\rm SSC}/\nu)$ decide the parameter dependence \citep{Sari_2001}. Ignoring the contribution from the logarithmic term, an analytical expression for dependencies of physical parameters can be obtained as $f_{\nu}^{\rm SSC} \propto E^{4/3} n_0^{1.467} \Gamma_0^{2.933} \epsilon_B^{0.8} \epsilon_e^{2.4}$. Here, we assumed the analytical expression of $t_{\rm dec}$.

For relatively higher $\epsilon_B$ values, the electrons are fast cooling at the peak time, and $\nu_c^{\rm ssc}$ goes below $500$~GeV. In this regime the parameter dependencies at $t_{\rm dec}$ are $f_{\nu}^{\rm SSC} \propto E^{2/3} n_0^{1/12} \Gamma_0^{-1/3} \epsilon_B^{-5/4}$. 

Therefore, the $500$~GeV flux is directly proportional to all parameters except $\epsilon_B$ in both slow and fast cooling regimes. A higher $\epsilon_B$ leads to a higher flux in the slow-cooling regime and a lower flux in the fast-cooling regime (for $500$~GeV $> \rm{min}(\nu_m^{\rm ssc}, \nu_c^{\rm ssc})$)

\subsection{Parameter space favouring TeV detections}

In order to fully explore the parameter space determining the lightcurve peak, we generated $500$~GeV light curves for a wide range of parameters and obtained the peak flux. We fixed $\epsilon_e = 0.02$ and $p=2.2$. We varied $10^{49} < E < 10^{51}$~ergs, $0.01 < n_0 < 100$~$cm^{-3}$, $ 50< \Gamma_0 <500$, and $10^{-7} < \epsilon_B <0.1$. In figures- \ref{E VS n} to \ref{n Vs epB}, we present the observable peak flux after EBL correction against different pairs of parameters to examine the ranges where the afterglow raises above CTA sensitivity. Left-hand side panels correspond to $z=0.5$ and right-hand side panels are for $z=0.1$.  

As we have seen in figures- \ref{E VS n}, \ref{E Vs G}, \ref{n_G_2}, higher $E, n_0,$ and $\Gamma_0$ lead to a higher flux. Considering flux for $z=0.1$ as a proxy for detectability, as EBL attenuation is not significant at that $z$, for $\epsilon_B = 10^{-3}$ and $\Gamma_0 = 300$, a burst of $E < 10^{51}$~erg is difficult to be detected unless $n_0 > 1$ (see figure-\ref{E VS n}). The limiting number density reaches $\sim 10$ and $\sim 100$ respectively for energy of $10^{50}$ and $10^{49}$~ergs. The numbers above correspond to an $\epsilon_B = 0.001$ and $\Gamma_0=300$. 

The behaviour with $\epsilon_B$ is, however different (see figures-\ref{E_epB_2} and \ref{n Vs epB}). During the slow cooling phase (which in turn corresponds to low $\epsilon_B$ values), the peak flux increases with $\epsilon_B$. At higher values of $\epsilon_B$ the electrons will be in the fast cooling phase. In this phase, if the observed frequency is above $\nu_c^{\rm ssc}$, the lightcurve peak is proportional to $\epsilon_B^{-5/4}$ (see section-\ref{Spectral energy}). So in the fast-cooling regime, a higher $\epsilon_B$ reduces the flux. This can be seen as a change in the direction of the contours in figure-\ref{E_epB_2} and figure-\ref{n Vs epB}. Because of this, the peak flux will not increase with $\epsilon_B$ beyond a certain threshold. At low $E$ values, therefore, the peak flux can not go beyond a certain limit figure-\ref{E_epB_2}. 

However, if $E$ or $n_0$ increases, the value of $\epsilon_B$ required for having a fast cooling phase reduces. This is because the magnetic field increases when $E$, $n_0$ or $\epsilon_B$ increases, forcing $\gamma_c$ to decrease (see figure-\ref{fig2}). The same can be observed in figure-\ref{E_epB_2} and figure-\ref{n Vs epB}, as the value at which $\epsilon_B$ contours turn direction increases as $E$ or $n_0$ decreases. Since $\gamma_m/\gamma_c \propto E^{1/3} n_0^{2/3}$, this drift is more visible w.r.t $n_0$ as in figure-\ref{E_epB_2}. 

It must be noted that these numbers are for an $\epsilon_e$ of $0.02$ and $p=2.2$. For the slow cooling phase, the peak flux is highly sensitive to $\epsilon_e$. The flux $f_{\nu}^{\rm SSC} \propto \epsilon_e^{2(p-1)}$ for $\nu > \nu_m^{\rm SSC}$, and increase in $\epsilon_e$ to $0.1$ will lead to an increase in flux by $50 - 100$ depending on the value of $p$. Flux in the fast cooling phase does not depend on $\nu_m^{\rm SSC}$ and is insensitive to $\epsilon_e$.

On the other hand, for a higher $\epsilon_e$ the fast cooling phase is more likely as discussed in section-\ref{Spectral energy}, and the above increase is applicable only for relatively lower values of $E, n_0,$ and $\Gamma_0$.

\section{Discussions}\label{Discussion}
\subsection{Bright TeV afterglows}

Since 2018, six Gamma-Ray Bursts are detected in TeV by Cherenkov telescopes (MAGIC, H.E.S.S, and LHAASO). In 2016, A $\sim 3 \sigma$ detection is reported for GRB160821B. Of these, GRB190114C, GRB190829A, and GRB221009A were exceptionally bright in TeV with $50, \sim 22,$ and $>100 \sigma$ detections respectively. The bursts are at redshifts of $0.424, 0.0785,$ and $0.151$ respectively. Out of these both GRB190114C and GRB221009A had isotropic equivalent energies $E_{\gamma, \rm iso}
> 10^{53}$~ergs in $\gamma$-rays. GRB190829A had a relatively low $E_{\gamma, \rm iso}$ of $\sim 10^{50}$~erg. Up to $18$~TeV photons were detected in GRB221009, which is the first detection of GRBs $>10$~TeV.

Our calculations show that at $z=0.1$, for $E = 10^{50}$~ergs, number densities of $10 - 100$~${\rm cm}^{-3}$ and $\Gamma_0$ of $100 - 300$ can lead to fluxes $\sim 10^{-11}{\rm erg}~{\rm cm^{-2}~{\rm s^{-1}}}$ at $\epsilon_B=0.001$. Therefore, the detection of TeV photons from GRB190829A is not surprising at $z=0.0785$.

From figure-\ref{SED_all}, for $z \sim 0.1$, the afterglow at the peak can be exceptionally bright in 500 GeV depending on other parameters. After an EBL absorption factor of about $0.6$, a flux as high as $10^{-10}{\rm erg}~{\rm cm}^{-2}~s^{-1}$ can be detected in $500$ GeV. For all the parameter combinations we presented here, at $10$ TeV the flux can reach $\sim 10^{-10} {\rm erg}~{\rm cm}^{-2}~{\rm s^{-1}}$ after EBL attenuation correction. For GRB221009A at $z=0.151$ \citep{de2022grb}, if all key parameters such as $E, n_0,$ and $\Gamma_0$ are high enough, it is very well possible to detect 10 s of TeV photons at $z \sim 0.1$. Therefore, the detection of 10 TeV photons from nearby GRBs are likely to continue with LHAASO \footnote{\url{http://english.ihep.cas.cn/lhaaso/}}.

\subsection{Counterparts of local Neutron Star mergers}
Due to the large field of view, CTA can have the potential to detect electromagnetic counterparts from neutron star mergers detected by AdvLIGO/Virgo \citep{2014MNRAS.443..738B}. MAGIC has reported a $3.5 \sigma$ detection for the LAT detected short GRB160829A \citep{Inoue:2020kkk} at the redshift of $0.424$. Short GRBs have relatively lower isotropic equivalent energies as opposed to long GRBs, and they are also likely to happen in low-density surroundings. From figure-\ref{E VS n}, low kinetic energy jets $ < 10^{50}$ launched in low-density medium ($n_0 < 1$) are unlikely to be detectable at $z \sim 0.1$. Therefore, TeV detections of nearby short GRBs are not very promising prospects. However, LIGO detection of nearer ($d_L \sim 200$~Mpc, $z \sim 0.05$) Neutron Star mergers, if associated with a relativistic jet launched towards the observer, the prospects of TeV detections will largely improve. Such a detection will have significant implications in understanding the magnetic field at the shock downstream and the number densities of the surrounding medium.

\section{Conclusions}\label{conclusion}
In this paper, we have explored the sub-TeV emission from GRB afterglows for a range of physical parameters. The optically thin SSC light curve peaks at the epoch of deceleration of the fireball. We focused on the parameter dependence of the peak flux in 500 GeV, considering that as a representative frequency. 

If shock acceleration efficiency reduces once the acceleration time scale is larger than the cooling timescale, the 500 GeV lightcurve is solely from the SSC process. We have neglected pair production optical depth and KN effects in calculating the TeV light curve. These can affect our results and reduce the flux by a factor of 10 to 100 for higher $n_0, \epsilon_B,$ and low $\Gamma_0$.

Depending on the physical parameters, particularly on $\epsilon_B$ and $\Gamma_0$, the electron population could be in the fast or slow cooling phase at the time of peak. High $\epsilon_B$ and low $\Gamma_0$ lead to a higher magnetic field and hence a fast cooling population. For an electron population in the slow cooling phase, the peak flux increases with $\epsilon_B$ but for a fast cooling population, peak flux decreases with $\epsilon_B$. 

In general, afterglows are TeV bright for higher $E$, $n_0$, and $\Gamma_0$. However, nearby bursts with energies $\sim 10^{50}$~ergs, such as GRB190829A, can be detected by Cherenkov telescopes if $n_0$ or $\Gamma_0$ is high enough.  Photons of energies as high as $18$~TeV were detected in GRB221009 at $z=0.15$. EBL attenuation at $z \sim 0.1$ can reduce the flux by a factor of 2. From our calculation, we see that if $E, n_0,$ and $\Gamma_0$ are favourable, the high energy photon detection rate by CTA will be promising. 

However, because of the lower energy and number densities, CTA detection of GRBs originating from merging neutron stars is not very likely. This will also affect LIGO counterparts from the nearby universe ($\sim 100$~Mpc) if the viewing angles are of the order of a few degrees.

\section*{Acknowledgements}
We thank the anonymous reviewer for his/her thoughtful comments, which have improved the quality of the manuscript.
T. Mondal acknowledges the support of the Prime Minister's Research Fellowship (\href{https://www.pmrf.in/}{PMRF}). T. Mondal like to thank Prof. Sonjoy Majumder, Dept. of Physics, IIT Kharagpur, whose constant guidance has been invaluable.
T. Mondal would also like to acknowledge S. Chakraborty, PhD Research Scholar at Dept of Physics, IIT Kharagpur, for his assistance with operating the \href{http://www.hpc.iitkgp.ac.in/}{Paramshakti Supercomputer facility} at IIT Kharagpur—a national supercomputing mission of the Government of India for providing High-Performance Computational resources. 
S. Pramanick acknowledges the support of \href{https://online-inspire.gov.in/}{DST-INSPIRE} Scholarship and Prime Minister's Research Fellowship (\href{https://www.pmrf.in/}{PMRF}).
L. Resmi acknowledges the Matrics grant MTR/2021/000830 of the Science and Engineering Research Board (SERB) of India.
D. Bose acknowledges the support of Ramanujan Fellowship-SB/S2/RJN-038/2017.
We use the publicly available package {\texttt JetSeT} to calculate the EBL attenuation factor.

\section*{Data Availability}
The numerical codes used in this study will be shared upon a reasonable request to the corresponding authors.

\begin{figure*}
\centering
\subfloat{\includegraphics[width = 0.5\textwidth]{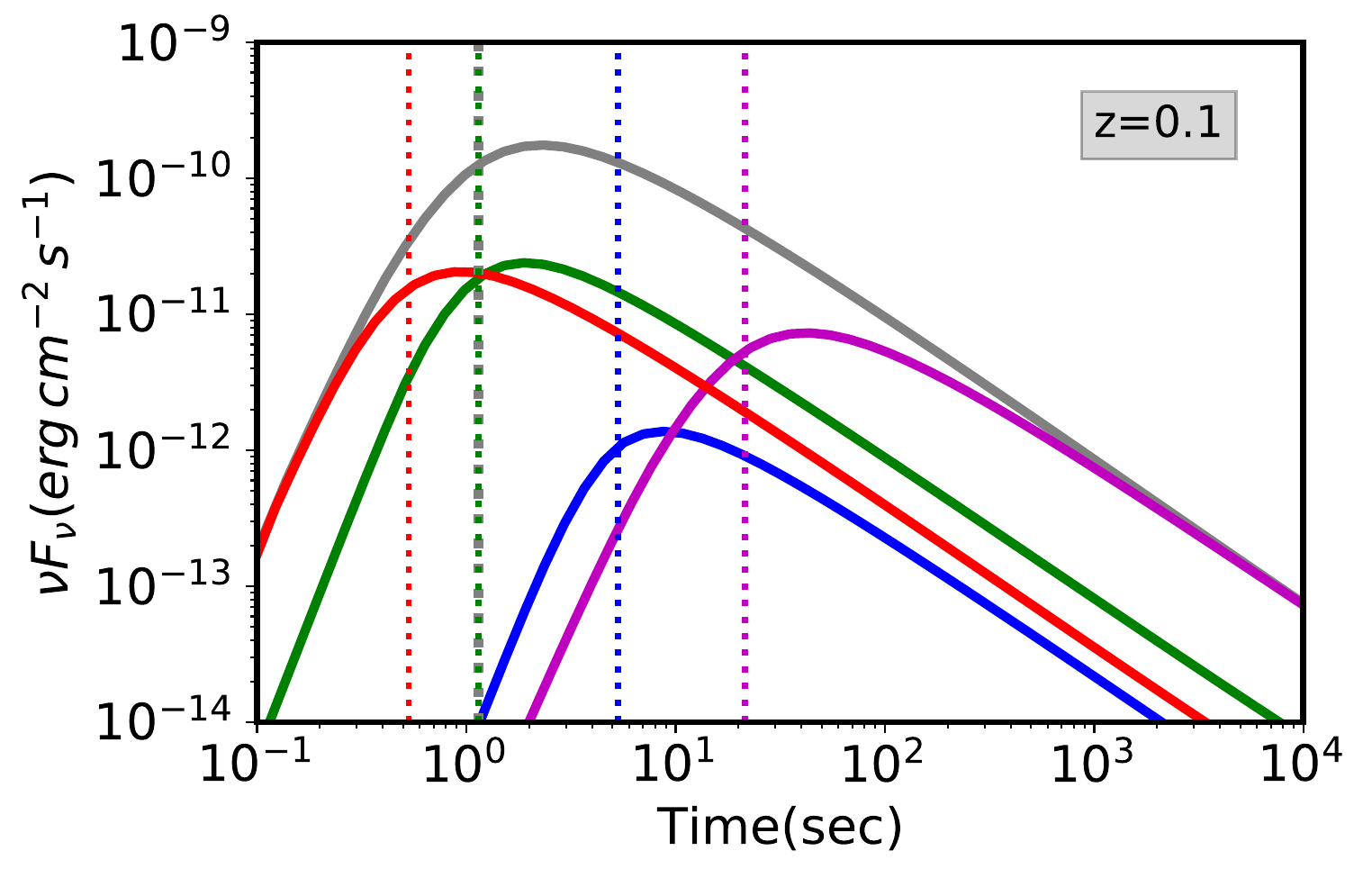}}
\subfloat{\includegraphics[width = 0.5\textwidth]{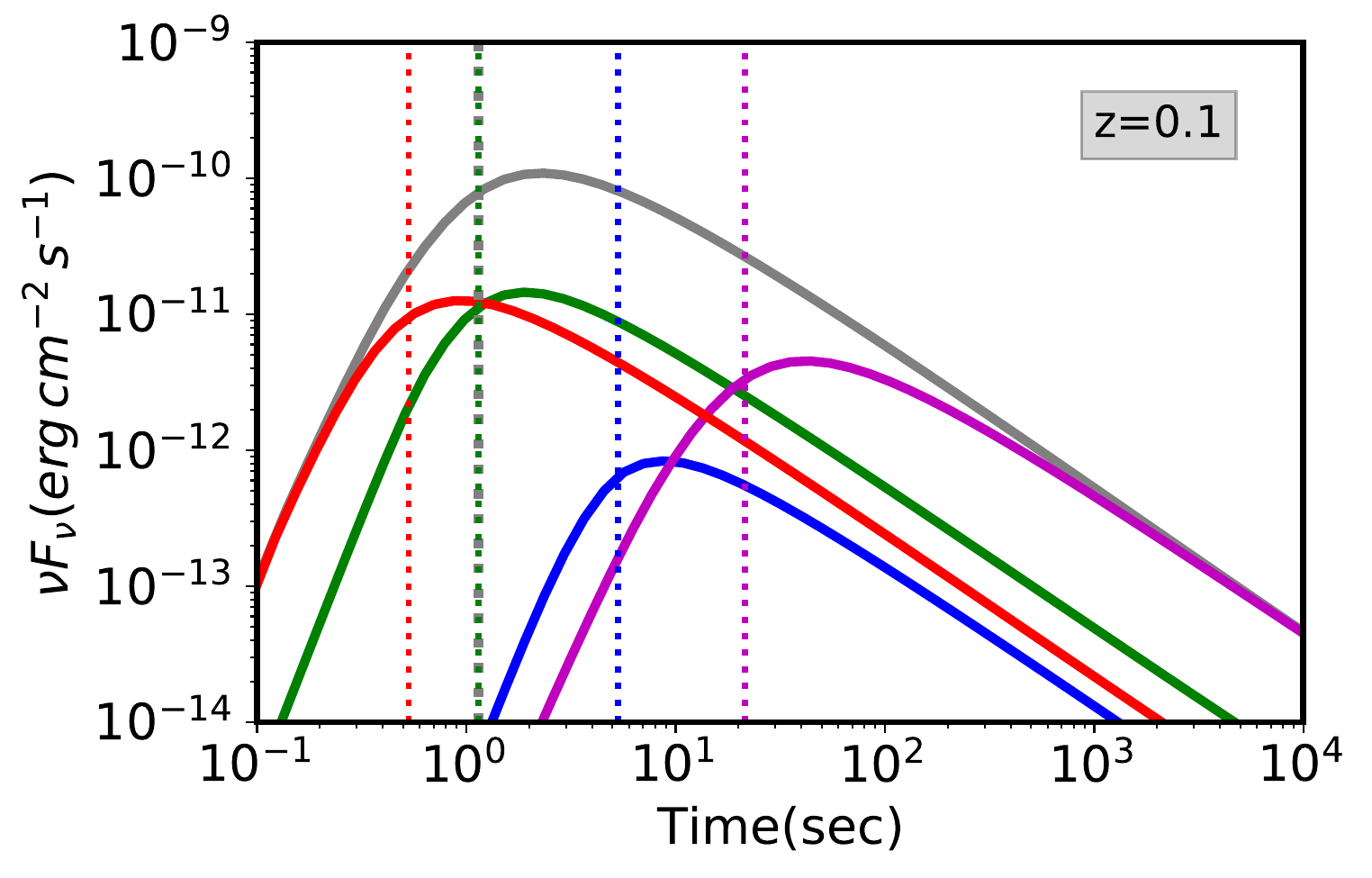}}

\caption{SSC Flux as a function of time is plotted for different combinations of microphysical parameters for redshift $z=0.1$, we set grey coloured lightcurve as the reference once $\left[E=10^{51} {\rm erg},\, \Gamma_{0}=300,\, n_{0}=10,\, \epsilon_{B}=10^{-3}\right]$. Blue, green, magenta and red coloured lines show the lightcurves for parameters $\left[E=10^{51} {\rm erg},\, \Gamma_{0}=300,\, n_{0}=0.1,\, \epsilon_{B}=10^{-3}\right]$, $\left[E=10^{51} {\rm erg},\, \Gamma_{0}=300,\, n_{0}=10,\, \epsilon_{B}=10^{-5}\right]$, $\left[E=10^{51} {\rm erg},\, \Gamma_{0}=100,\, n_{0}=10,\, \epsilon_{B}=10^{-3}\right]$, $\left[E=10^{50} {\rm erg}, \,\Gamma_{0}=300,\, n_{0}=10,\, \epsilon_{B}=10^{-3}\right]$ respectively, at some fixed parameters $p=2.2,\, \epsilon_{e}=0.02$ at redshift $z=0.1$. The lightcurve plot of the left-side panel represents SSC Flux without EBL correction, and the right-side panel depicts the SSC flux with  EBL correction  by EBL Dominguez factor $= 0.6013$ at $500$ GeV.}
\label{LC_1_all}
\end{figure*}

\begin{figure*}
\subfloat{\includegraphics[width = 0.5\textwidth]{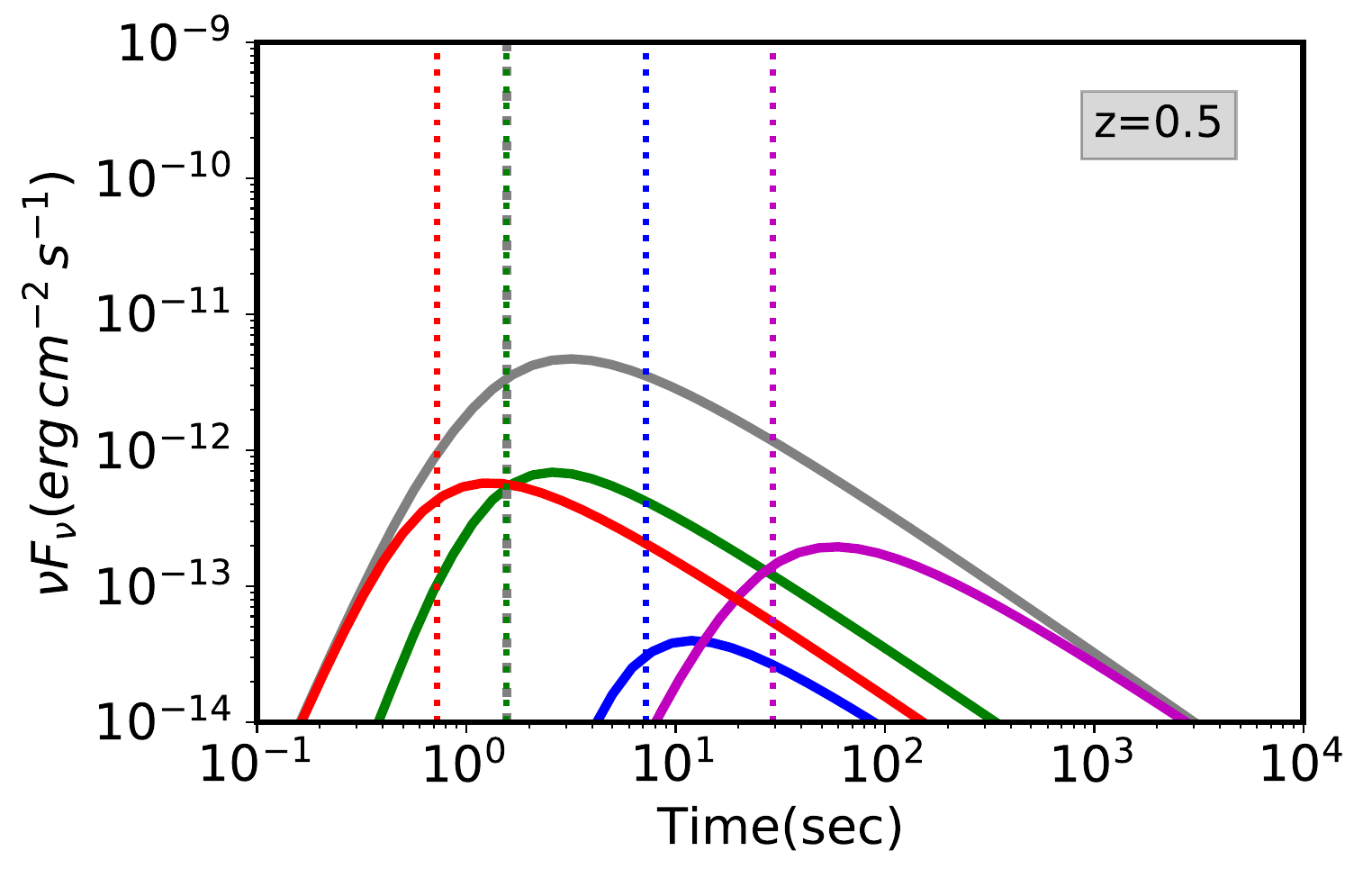}}
\subfloat{\includegraphics[width = 0.5\textwidth]{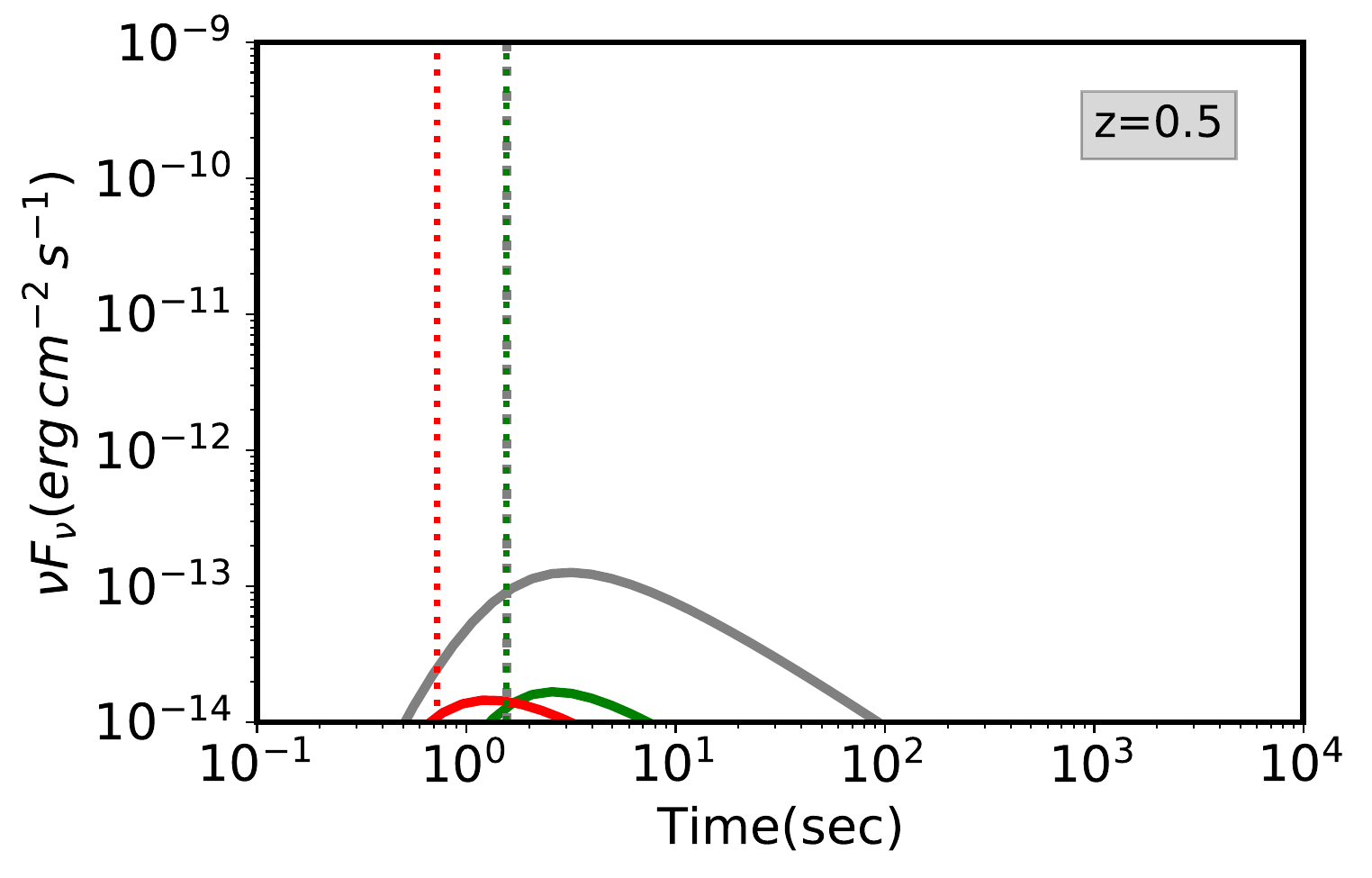}}

\caption{SSC Flux as a function of time is plotted for the same combination of microphysical parameters with a similar colour scheme as mentioned in figure-\ref{LC_1_all}, at the same fixed parameters with different redshift value at $z=0.5$. The lightcurve plot of the left-side panel represents SSC Flux without EBL correction, and the right-side panel depicts the SSC flux with EBL correction  by EBL Dominguez factor $= 0.0233$ at $500$ GeV.}
\label{LC_2_all}
\end{figure*}

\begin{figure*}
{\includegraphics[width = 0.5\textwidth,left]{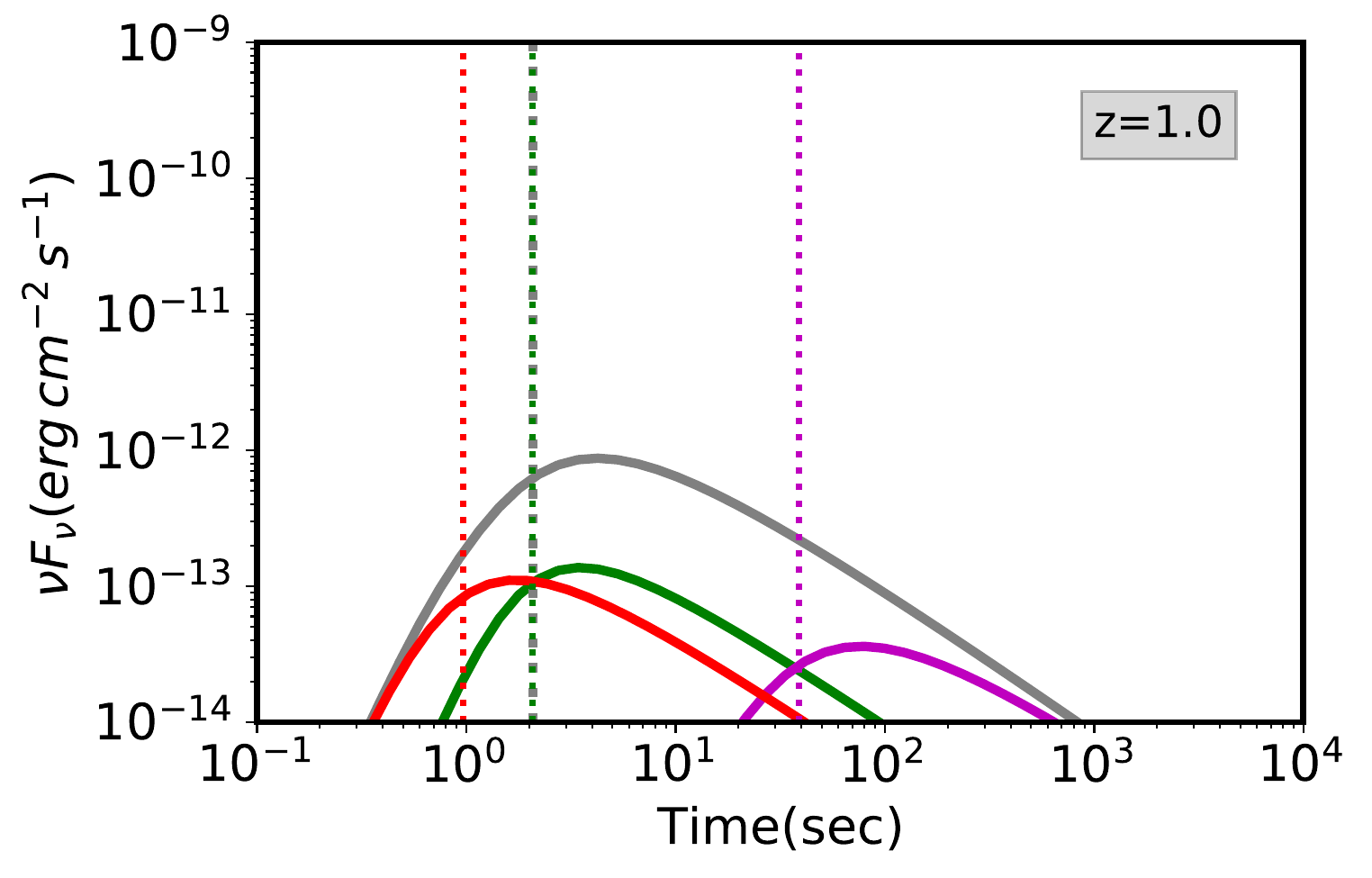}}

\caption{The lightcurve represents SSC Flux without EBL correction. With EBL Dominguez factor $=0.00012$, at $500$ GeV for $z=1.0$, the EBL corrected SSC flux is going below the value of $10^{-14}{\rm erg}\, {\rm cm}^{-2}\, s^{-1}$, which is beyond the reach of CTA. Hence, the EBL corrected plot is discarded.}
\label{LC_3_all}
\end{figure*}


\begin{figure*}
    \centering
    \includegraphics[width = 0.98\textwidth]{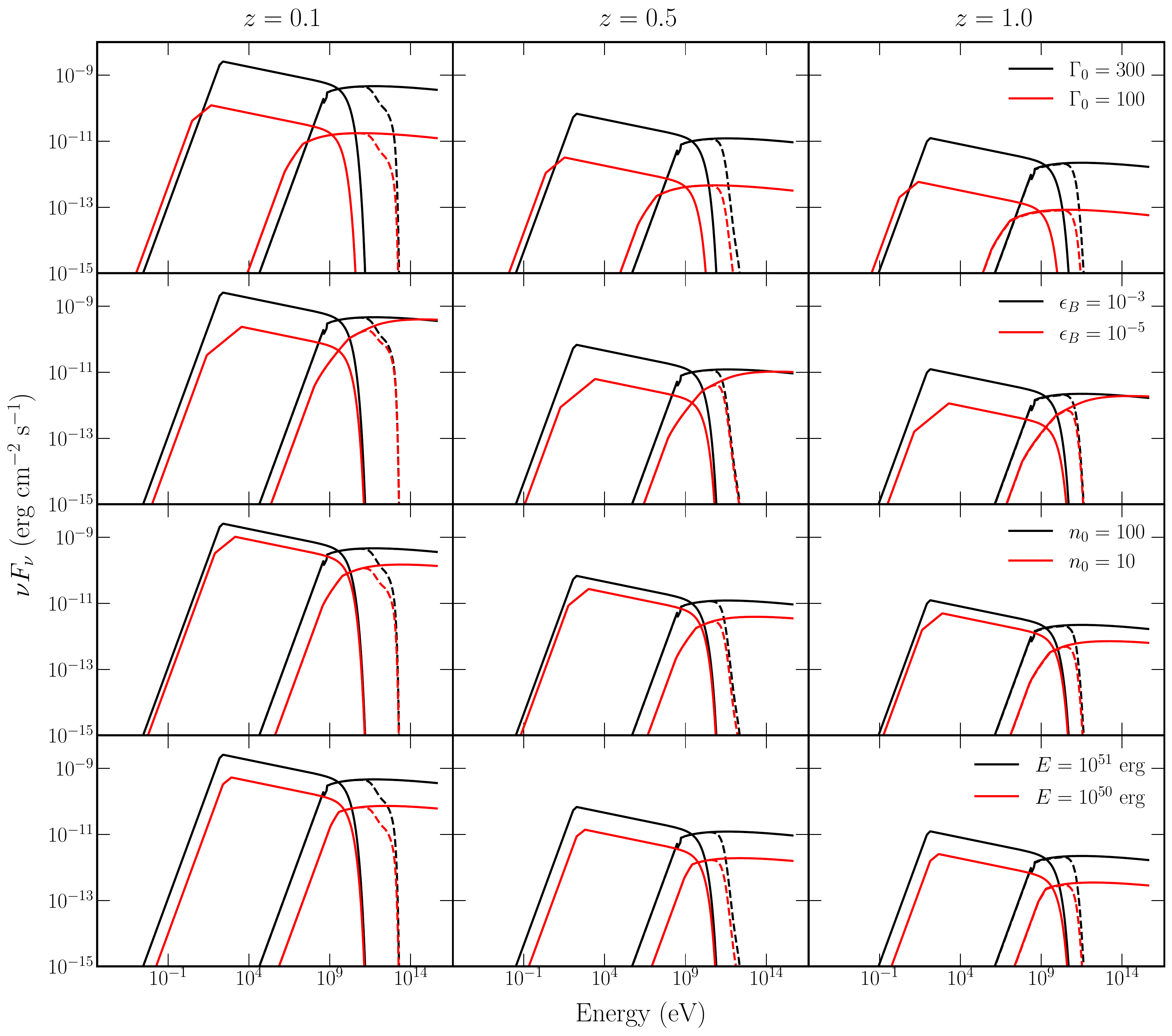}
    \caption{This shows the SEDs for different parameters and redshifts. From left to right column redshift values are $z=0.1, \: 0.5$ and $1.0$. The first row depicts the plots with fixed $E = 10^{51}$ erg, $n_{0}=100$, $\epsilon_{B}=10^{-5}$. For this row the black line represnets $\Gamma_{0} =300$ and red line is for $\Gamma_{0} = 100$.  For the second row,  $E=10^{51}$ erg, $n_{0}=100$, $\Gamma_{0} =300$ are fixed, $\epsilon_{B}=10^{-3}$ is the black line and $\epsilon_{B}=10^{-5}$ is the red line. In third row,  $E=10^{51}$ erg, $\epsilon_{B}=10^{-3}$, $\Gamma_{0} =300$ are fixed, $n_{0}=100$ and $n_{0}=10$ are black and red lines respectively.  For fourth row, $n_{0}=100$, $\epsilon_{B}=10^{-3}$, $\Gamma_{0} =300$ are fixed, $E=10^{51}$ erg and $E=10^{50}$ erg  correspond to black and red lines respectively. All the solid lines represent flux without EBL corrections, whereas, the dashed lines represent the flux after EBL corrections.}
    \label{SED_all}
\end{figure*}

\begin{figure*}
\centering
\subfloat{\includegraphics[width = 0.5\textwidth]{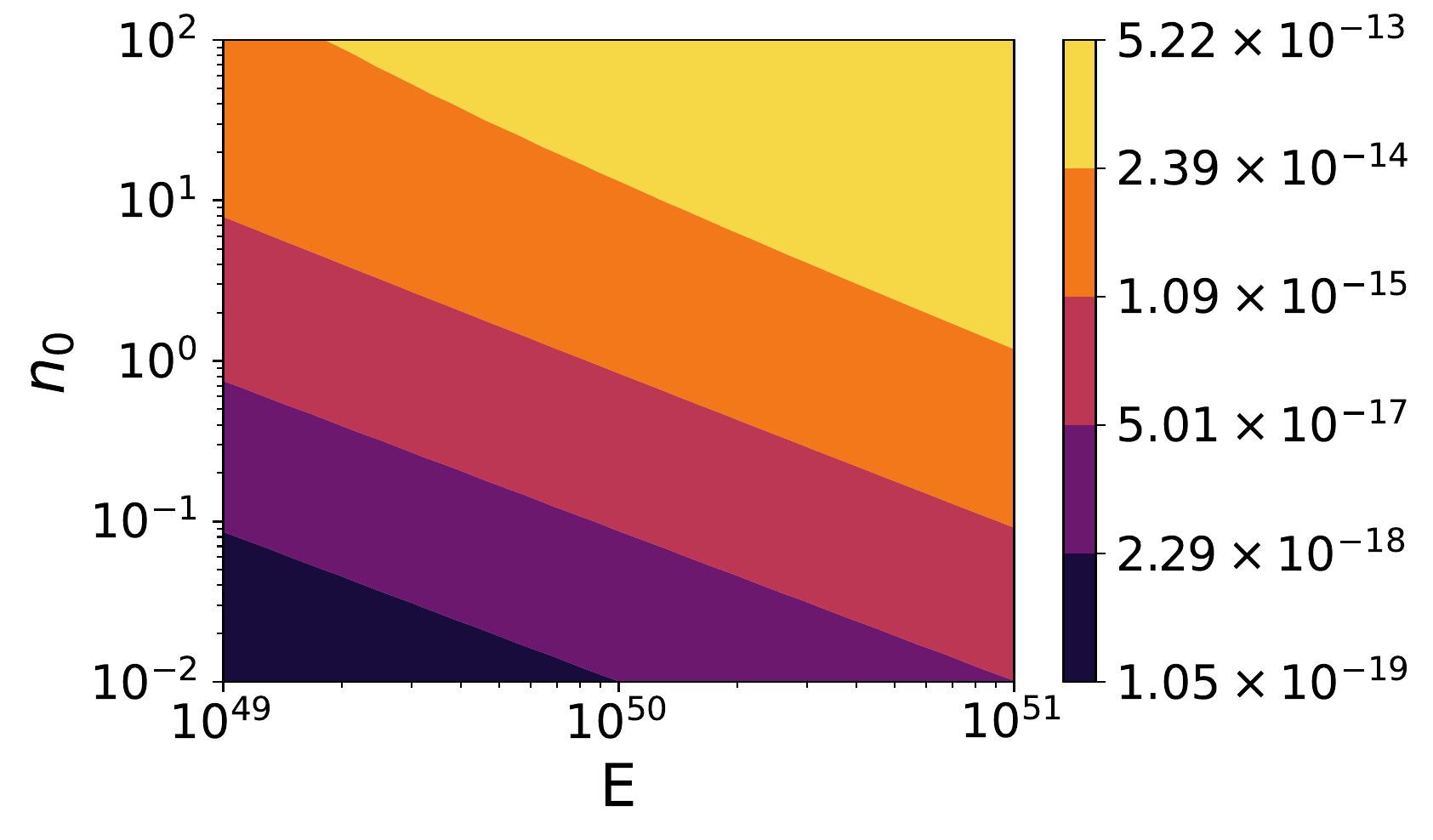}}
\subfloat{\includegraphics[width = 0.5\textwidth]{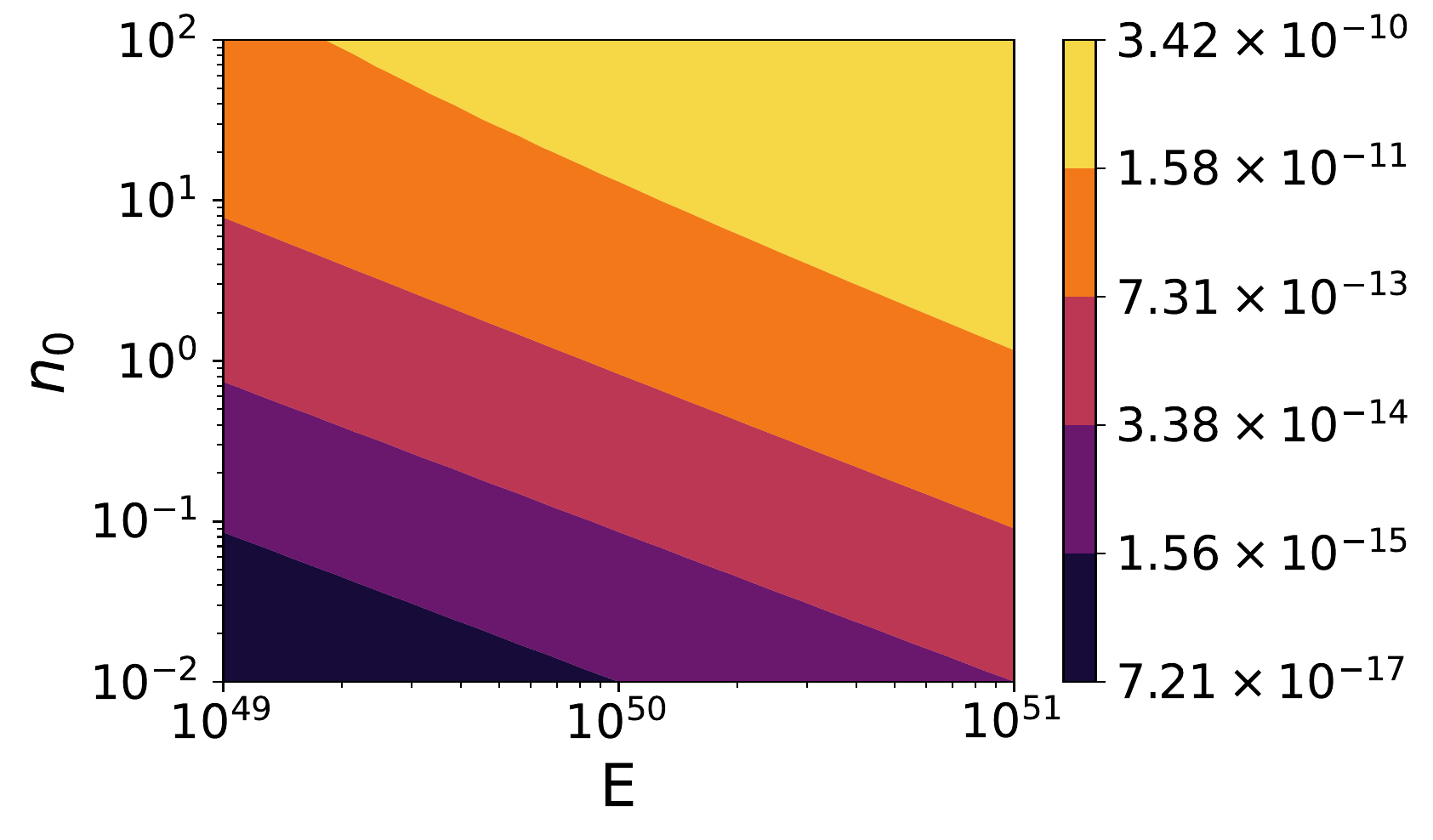}}

\caption{These plots depict the EBL corrected SSC peak flux for different $E$ and $n_{0}$ values, with other parameters fixed as $\epsilon_{e}=0.02$,  $\epsilon_{B}=0.001$, $\Gamma_{0}= 300$. The left and right panels are for redshift $z=0.5$ and $z=0.1$ respectively.}
\label{E VS n}
\end{figure*}

\begin{figure*}
\centering
\subfloat{\includegraphics[width = 0.5\textwidth]{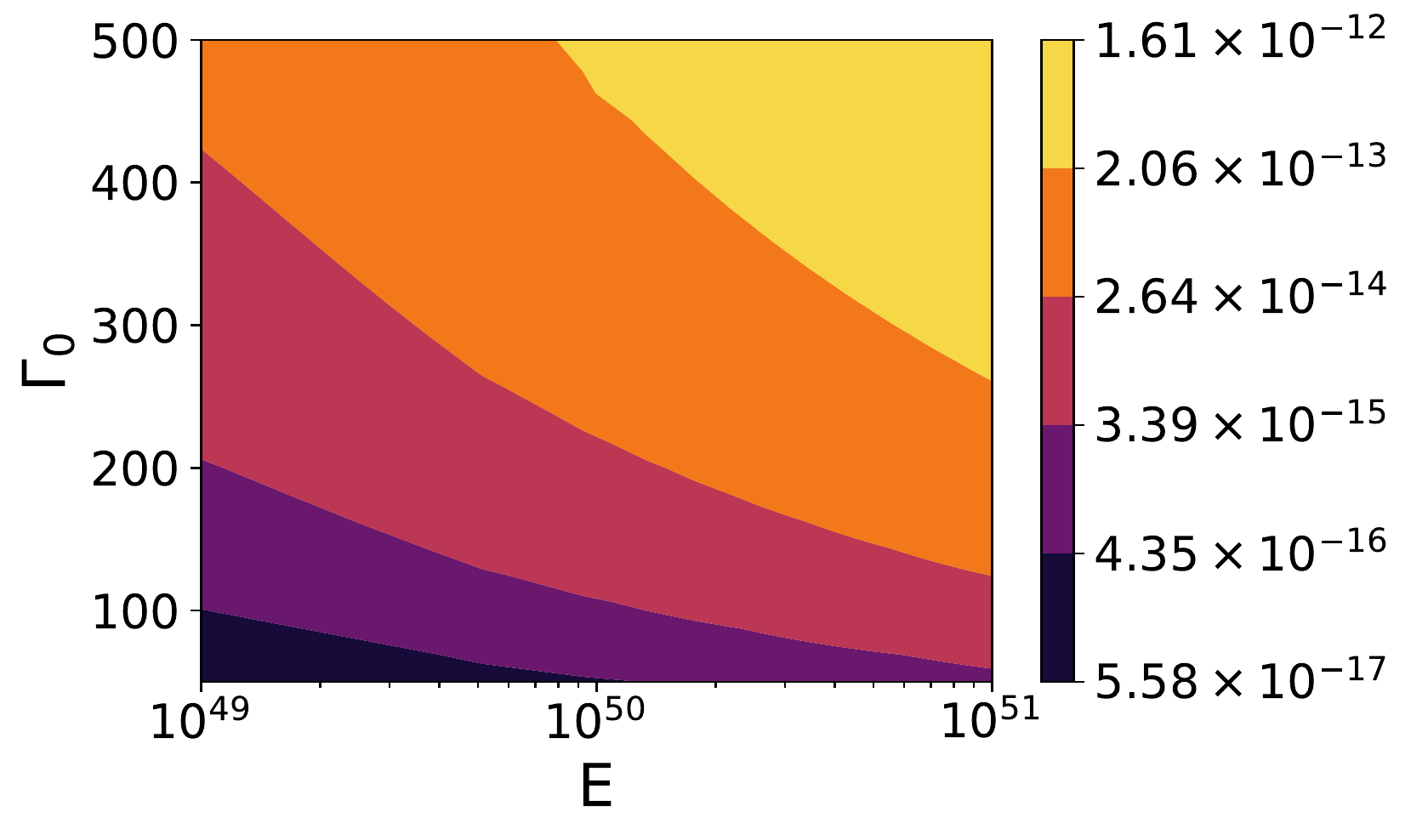}}
\subfloat{\includegraphics[width = 0.5\textwidth]{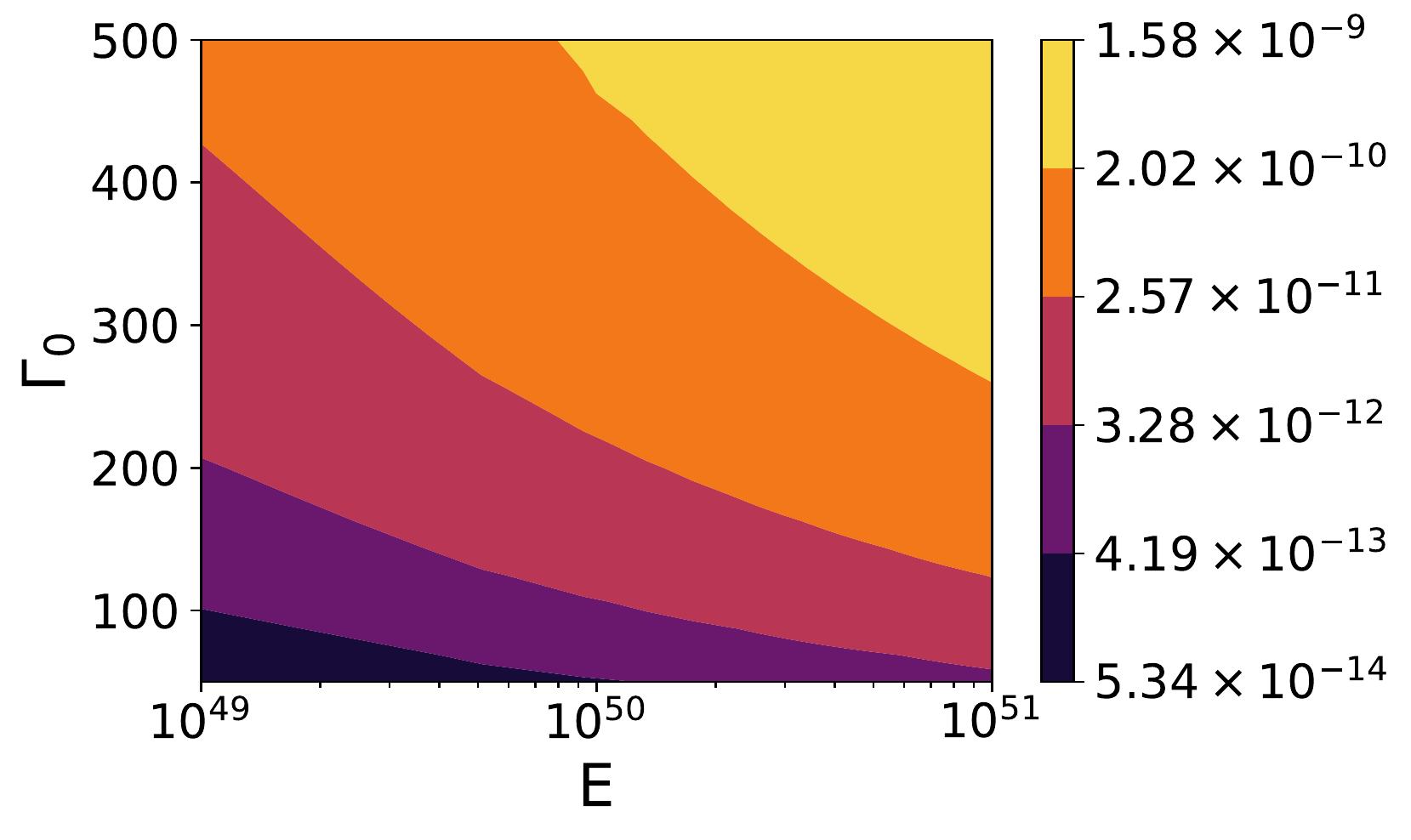}}
\caption{These plots show the EBL corrected SSC peak flux for different $E$ and $\Gamma_{0}$ values, with other parameters fixed as $\epsilon_{e}=0.02$,  $\epsilon_{B}=0.001$, $n_{0}=100$. The left panel and right panels are for redshift $z=0.5$ and $z=0.1$, respectively.}
\label{E Vs G}
\end{figure*}

\begin{figure*}
\centering
\subfloat{\includegraphics[width = 0.5\textwidth]{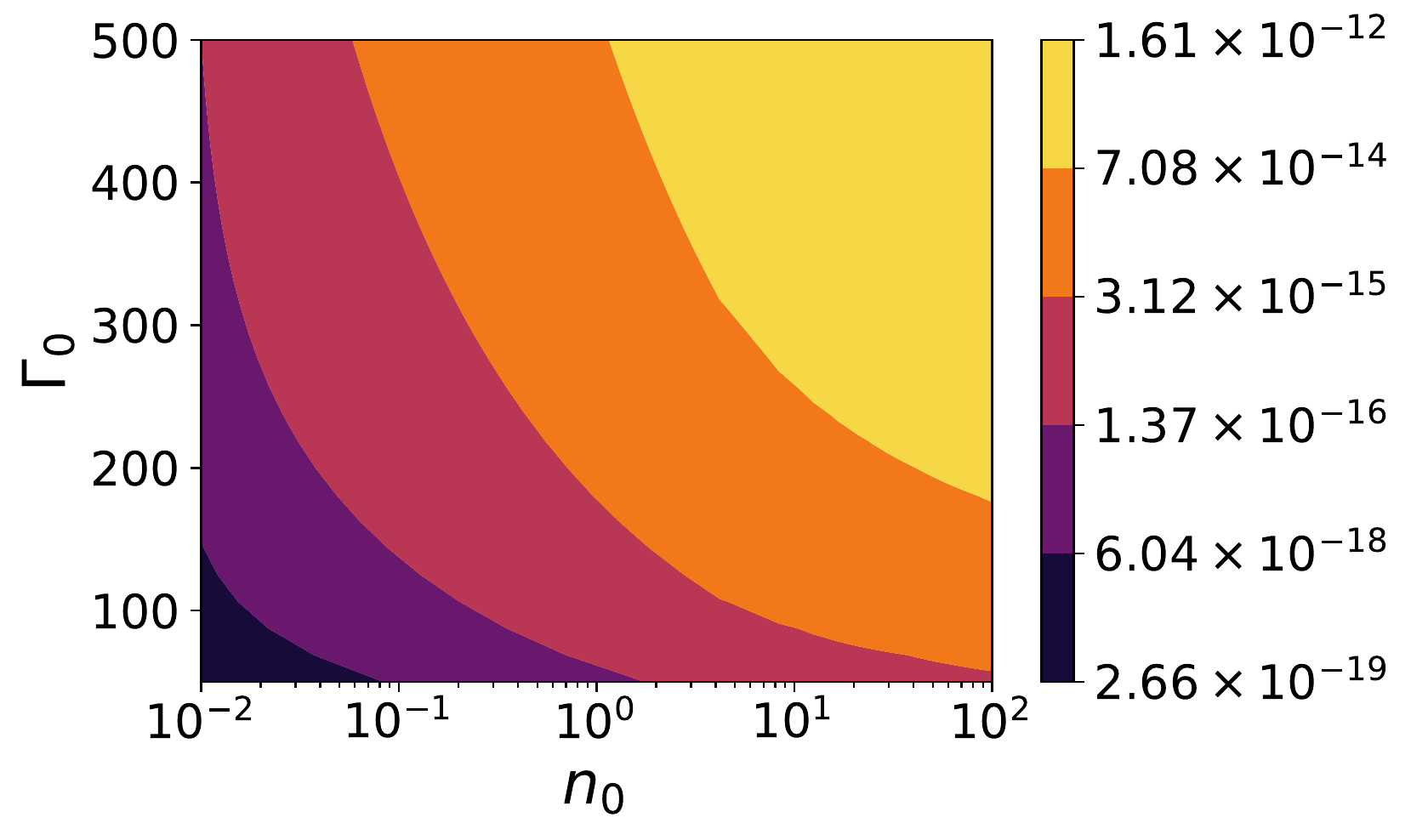}}
\subfloat{\includegraphics[width = 0.5\textwidth]{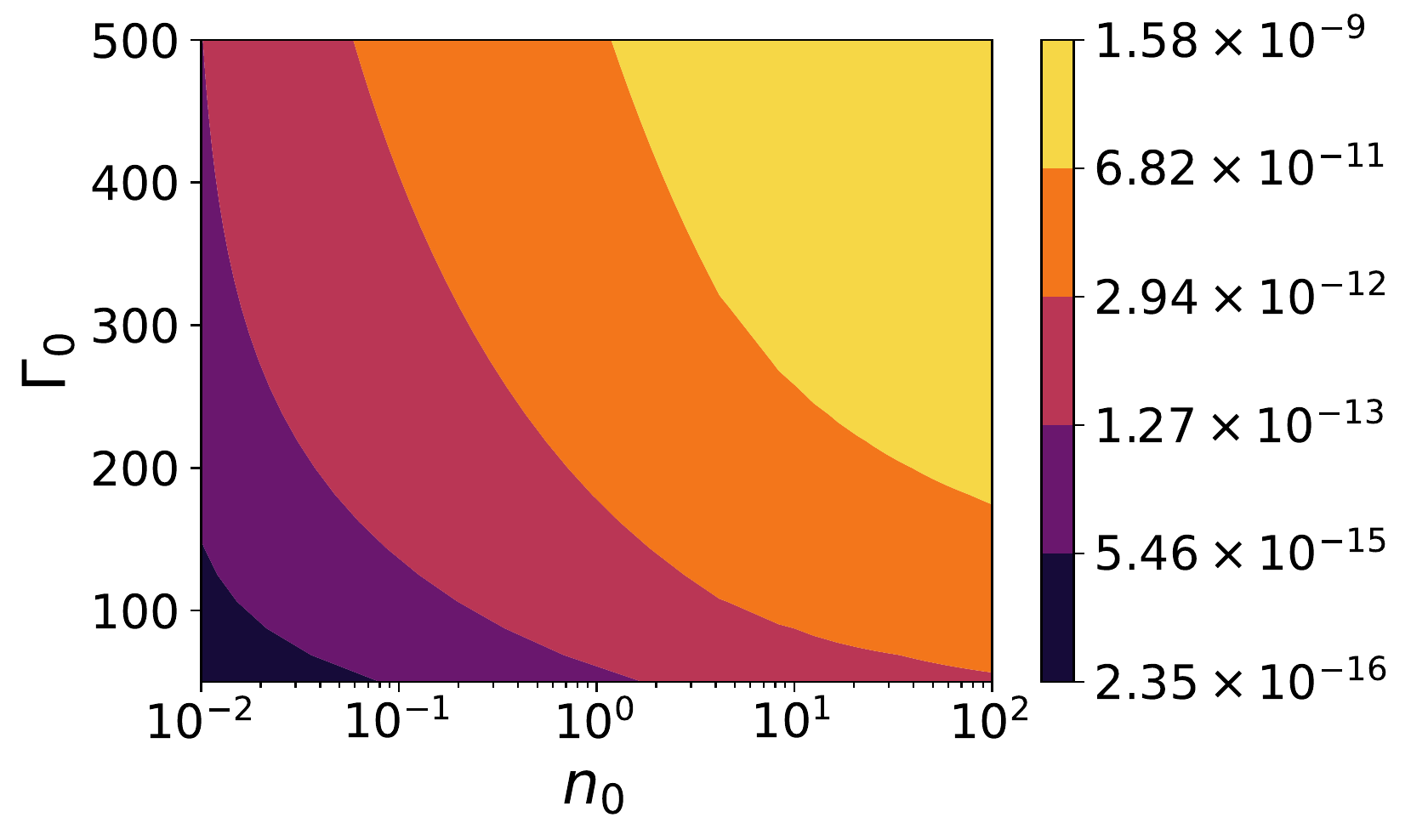}}
\caption{These plots show the EBL corrected SSC peak flux for different $n_{0}$ and $\Gamma_{0}$ values, with other parameters fixed as $\epsilon_{e}=0.02$,  $\epsilon_{B}=0.001$, $E=10^{51}$ erg. The left and right panels are for redshift $z=0.5$ and $z=0.1$, respectively.}
\label{n_G_2}
\end{figure*}

\begin{figure*}
\centering
\subfloat{\includegraphics[width = 0.5\textwidth]{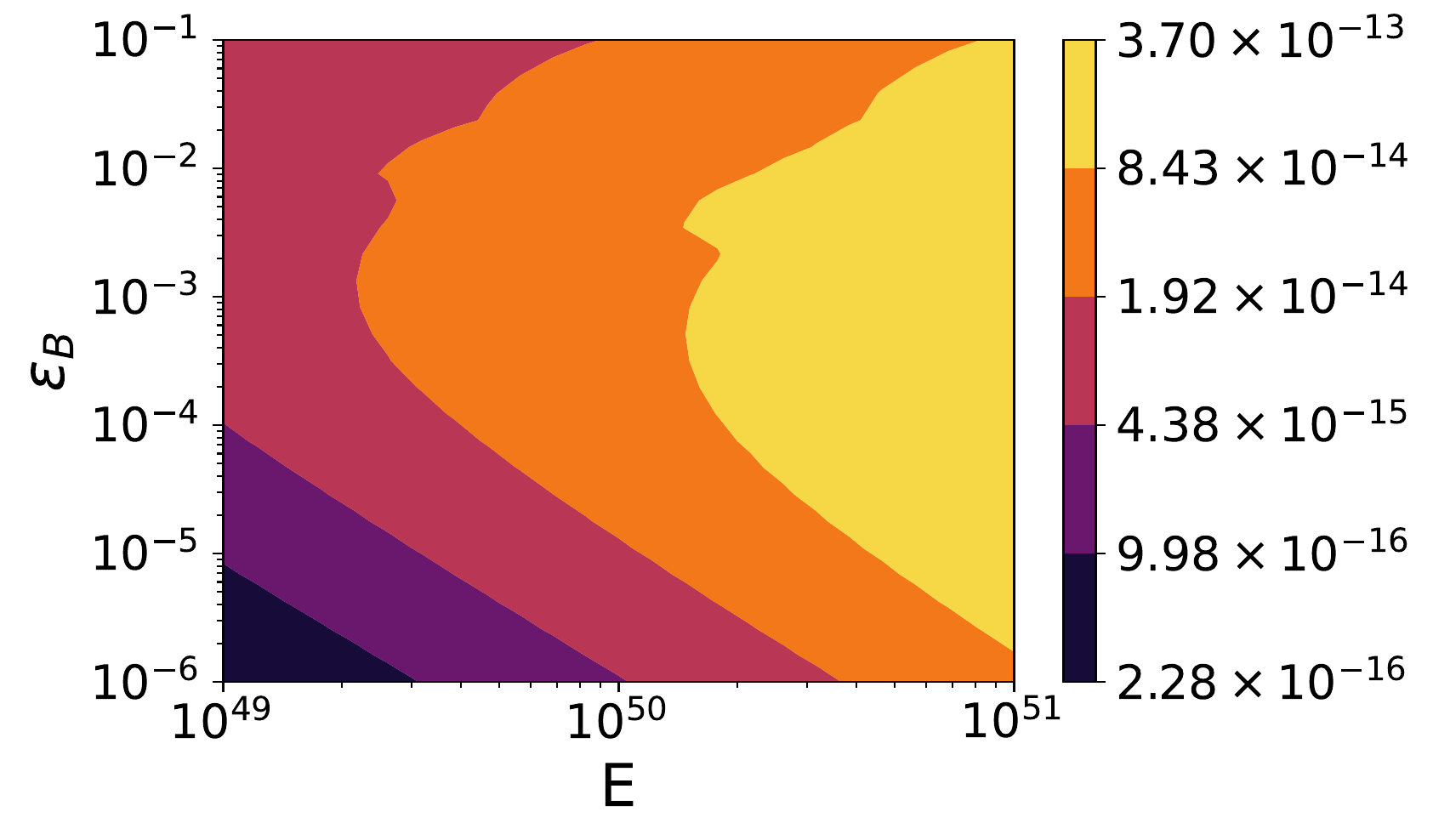}}        
\subfloat{\includegraphics[width = 0.5\textwidth]{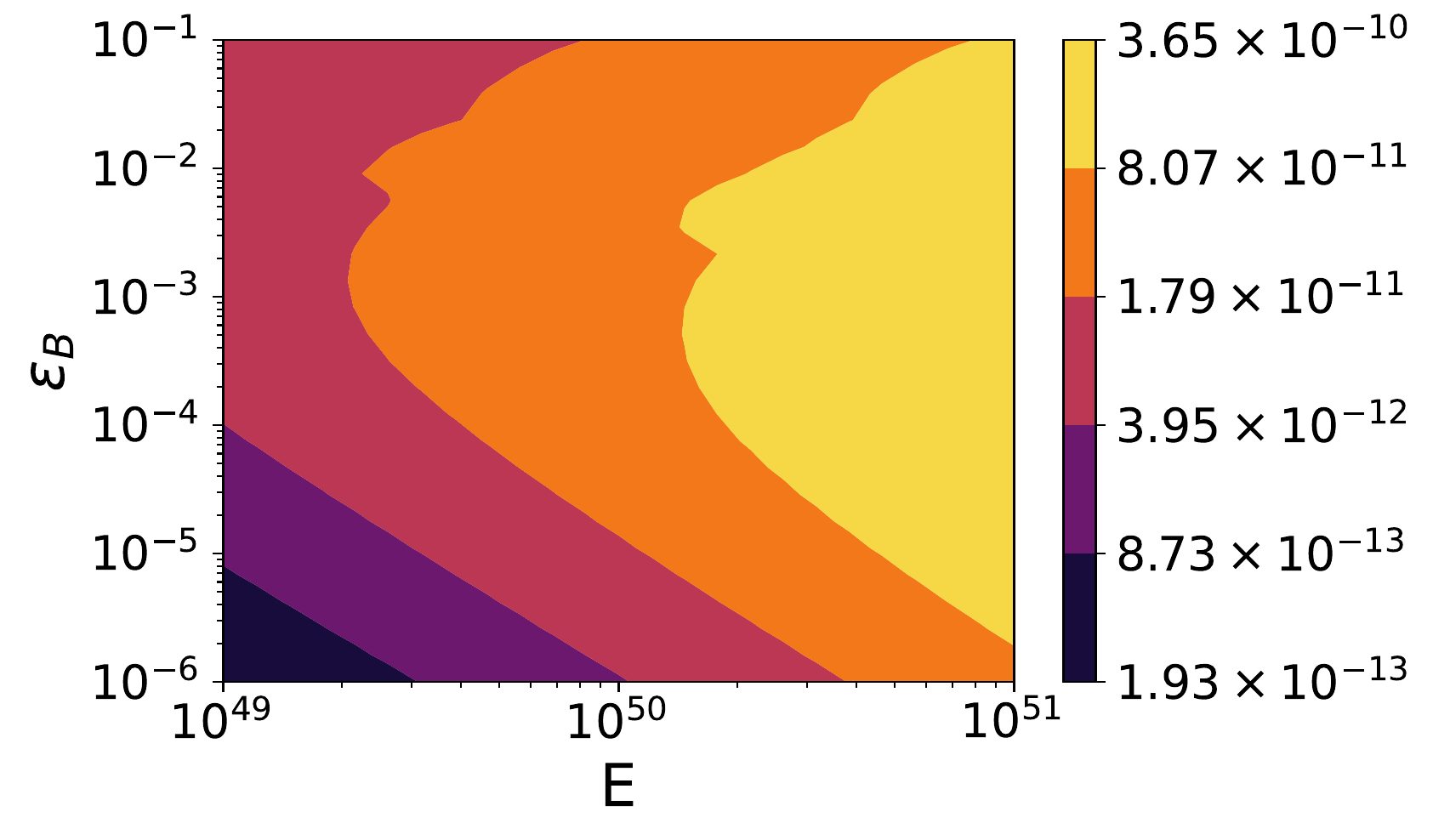}}
\caption{These plots show the EBL corrected SSC peak flux for different $E$ and $\epsilon_{B}$ values, with other parameters fixed as $\epsilon_{e}=0.02$,  $n_{0}=100$, $\Gamma_{0}=300$. The left and right panels are for redshift $z=0.5$ and $z=0.1$, respectively.}
\label{E_epB_2}
\end{figure*}

\begin{figure*}
\centering
\subfloat{\includegraphics[width = 0.5\textwidth]{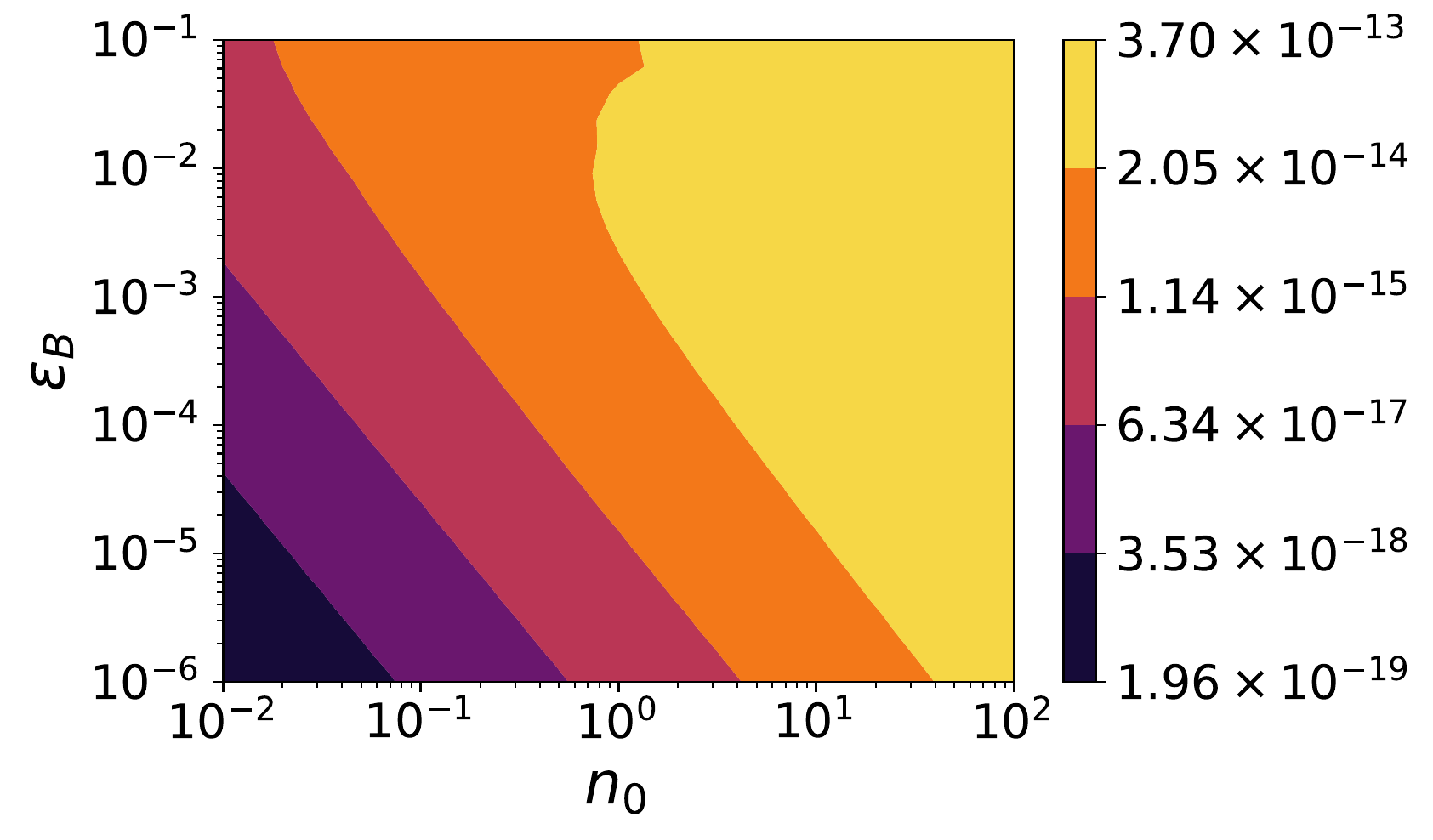}} 
\subfloat{\includegraphics[width = 0.5\textwidth]{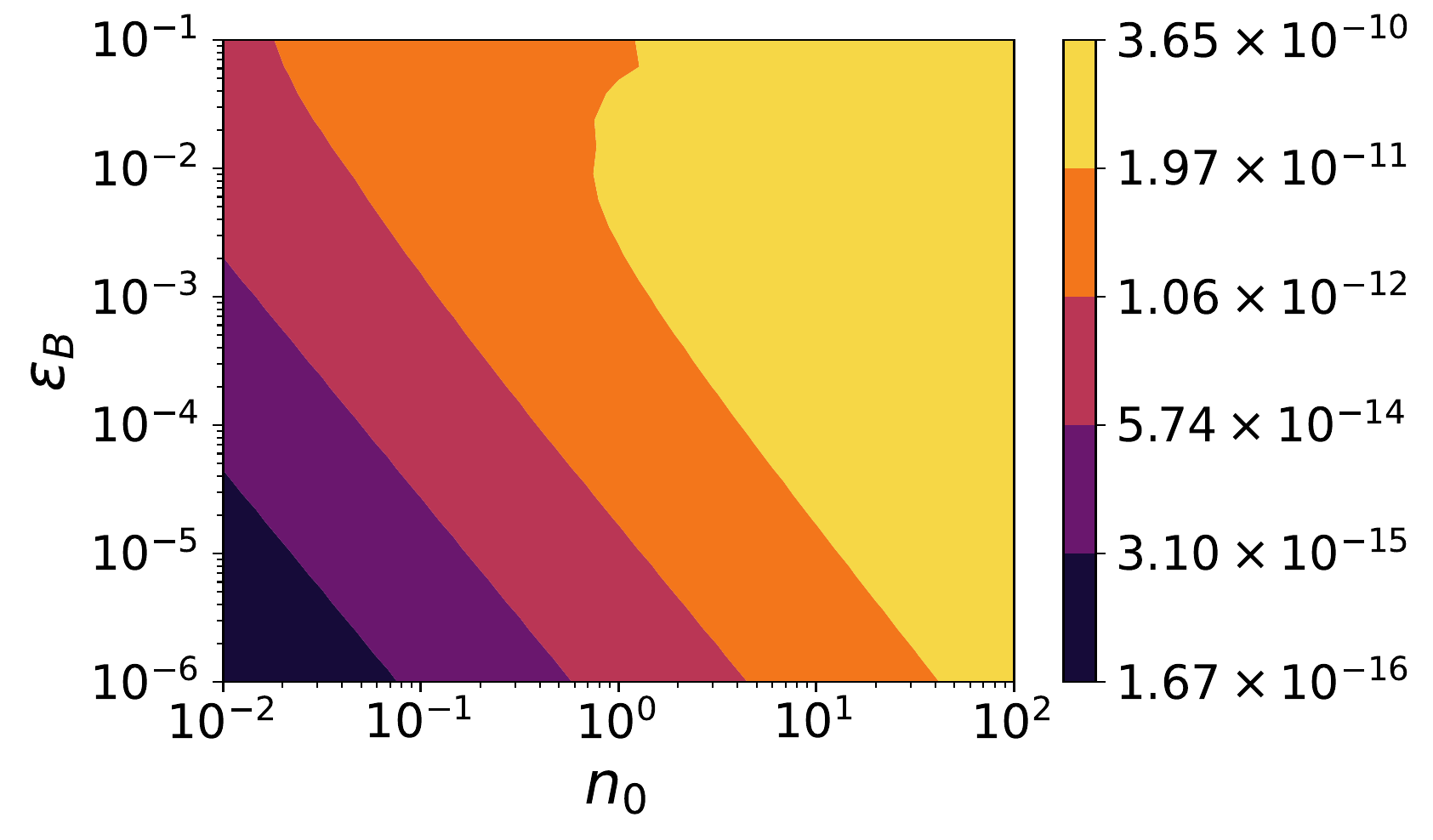}} 
\caption{These plots show the EBL corrected SSC peak flux for different $n_{0}$ and $\epsilon_{B}$ values, with other parameters fixed as $\epsilon_{e}=0.02$,  $\Gamma_{0}=300$, $E=10^{51}$ erg. The left and right panels are for redshift $z=0.5$ and $z=0.1$, respectively.}
\label{n Vs epB}
\end{figure*}

\bibliographystyle{mnras}
\bibliography{GRB}

\appendix

\bsp	
\label{lastpage}
\end{document}